\documentclass[fleqn,usenatbib]{mnras}

\usepackage{newtxtext,newtxmath}

\usepackage[T1]{fontenc}
\usepackage{ae,aecompl}

\usepackage{graphicx}	
\usepackage{amsmath}	
\usepackage{amssymb}	
\usepackage{mathtools}
\usepackage{hyperref}
\usepackage{xspace}	
\usepackage{euscript}
\usepackage{xcolor}
\usepackage{listings}
\usepackage{textcomp}

\newcommand{\edits}[1]{{#1}}
\newcommand{\gaia}{{\it Gaia}\xspace}

\definecolor{codegreen}{rgb}{0,0.6,0}
\definecolor{codegray}{rgb}{0.5,0.5,0.5}
\definecolor{codepurple}{rgb}{0.58,0,0.82}
\definecolor{backcolour}{rgb}{0.95,0.95,0.92}

\lstdefinestyle{python}{
    backgroundcolor=\color{backcolour},   
    commentstyle=\color{codegreen},
    keywordstyle=\color{magenta},
    numberstyle=\tiny\color{codegray},
    stringstyle=\color{codepurple},
    basicstyle=\ttfamily\footnotesize,
    breakatwhitespace=false,         
    breaklines=true,                 
    captionpos=b,                    
    keepspaces=true,                 
    numbers=left,                    
    numbersep=5pt,                  
    showspaces=false,                
    showstringspaces=false,
    showtabs=false,                  
    tabsize=2,
    upquote=true
}

\title[What are the odds that a star is missing?]{Completeness of the \textit{Gaia}-verse II: what are the odds that a star is missing from \textit{Gaia} DR2?}

\author[D. Boubert and A. Everall]{
	Douglas Boubert$^{1}$\thanks{E-mail: douglas.boubert@magd.ox.ac.uk}
	and Andrew Everall$^{2}$
	\\
	$^{1}$Magdalen College, University of Oxford, High Street, Oxford OX1 4AU, UK\\
	$^{2}$Institute of Astronomy, University of Cambridge, Madingley Road, Cambridge CB3 0HA, UK\\
}

\date{Accepted XXX. Received YYY; in original form ZZZ}

\pubyear{}

\begin{document}
\label{firstpage}
\pagerange{\pageref{firstpage}--\pageref{lastpage}}
\maketitle

\begin{abstract}
	The second data release of the \gaia mission contained astrometry and photometry for an incredible 1,692,919,135 sources, but how many sources did \gaia miss and where do they lie on the sky? \edits{The answer to this question will be crucial for any astronomer attempting to map the Milky Way with \gaia DR2.} We infer the completeness of \gaia DR2 by \edits{exploiting the fact that it only contains sources with at least five astrometric detections}. \edits{The odds that a source achieves those five detections depends on both the number of observations and the probability that an observation of that source results in a detection.} We predict the number of times that each source was observed by \gaia and assume that the probability of detection is either a function of magnitude or a distribution as a function of magnitude. We fit both these models to the 1.7 billion stars of \gaia DR2, and thus are able to robustly predict the completeness of \gaia across the sky as a function of magnitude. We extend our selection function to account for crowding in dense regions of the sky, and show that this is vitally important, particularly in the \edits{Galactic bulge and the Large and Small Magellanic Clouds}. \edits{We find that the magnitude limit at which \gaia is still 99\% complete varies over the sky from $G=18.9$ to $21.3$.}  We have created a new \textsc{Python} package \textsc{selectionfunctions} (\url{https://github.com/gaiaverse/selectionfunctions}) which provides easy access to our selection functions.
\end{abstract}




\section{Introduction}

The \gaia mission \citep{Gaia2016,Gaia2018} will give us a many-dimensional astrometric, photometric and spectroscopic perspective on the stars of the Milky Way. Already with the preliminary second data release (DR2), \gaia has provided astrometric positions and broad-band $G$ photometry for 1,692,919,135 sources. \edits{This begs the question, however: which stars are missing from \gaia DR2 and where do they lie on the sky?}

In this second work in our series investigating the completeness of the \gaia-verse, we quantify the completeness of the stars with positions $(\alpha,\delta)$ and $G$ magnitudes in \gaia DR2 (i.e. all 1,692,919,135 sources included in \edits{the DR2 source catalogue}) through a selection function. \edits{Such a selection function would be} immediately useful for those using star count overdensities to look for stellar structures in the Galaxy \edits{or for those looking to map the distribution of stars in the disk}. Furthermore, it is a necessary foundation for the selection functions of the more commonly used \gaia DR2 subsets that have proper motions and parallaxes, colour photometry, \edits{variable star classifications}, or radial velocities.

The simplest way to calculate a catalogue's selection function is to count the fraction of stars in another more complete catalogue that are missing; for example, \citet{Bovy2014} calculated the selection function of the spectroscopic APOGEE Red Clump Catalog in comparison to the photometric 2MASS survey. The APOGEE survey selected stars for observation from the 2MASS source catalogue \citep{Zasowski2013} and so 2MASS is guaranteed to be a superset of the stars observed by APOGEE. This approach is attractive because it is easy to understand and implement, and does not require any knowledge about the instrumentation or pipeline. However, it does require that the comparison catalogue is truly complete for the magnitude range of the catalogue of interest, and so cannot be used to compute the \gaia selection function, because there is no more complete catalogue for \gaia to be compared against. At the other extreme, if we have perfect knowledge of the instrumentation and pipeline that lead to a catalogue, then the selection function of that catalogue can be directly computed. We adopt a hybrid approach in this work: by using a small amount of knowledge about \gaia (the spinning-and-precessing scanning law) we are able to compute an empirically-driven selection function.

The key and entirely novel idea behind this paper is that the selection function of \gaia DR2 is approximately that the number of times that \gaia detected the source as it crossed the field-of-view was greater than four, as motivated in the preamble of Sec. \ref{sec:methodology}. We use the term `detections' to mean those occasions where the source transiting the \gaia field-of-view resulted in an astrometric measurement that contributed to the source's astrometric solution. \edits{There are many reasons why an observation of a source might not result in such a detection, as discussed later in the text}. \edits{The} number of detections is termed \textsc{astrometric\_matched\_observations} in the \gaia terminology and this number is given for every star in \gaia DR2.

In the first part of this paper we assume that the probability that a star is detected on each observation is solely a function of brightness, and thus by modelling the \textsc{astrometric\_matched\_observations} of all the stars in \gaia DR2 we deduce a first-order selection function for the entire catalogue. We describe our methodology in Sec. \ref{sec:methodology} and present our inferred selection function in Sec. \ref{sec:results}, including a map of the magnitude limit of 99\% completeness in the top panel of Fig. \ref{fig:completenessmaps}.

In the second part of this paper, we extend our selection function to account for crowding. \gaia can only \edits{simultaneously track} $1,050,000\;\mathrm{sources}\;\mathrm{deg}^{-2}$ \citep{Gaia2016} and will choose to track brighter stars ahead of fainter stars, and thus crowding acts as a second-order effect which limits the completeness of \gaia with respect to faint stars in dense regions of the sky. The effect of crowding is vitally important in the Galactic bulge and the Large and Small Magellanic Clouds. We describe the adaptation of our selection function to account for crowding in Sec. \ref{sec:crowding}, and show a revised map of the magnitude limit of 99\% completeness in the bottom panel of Fig. \ref{fig:completenessmaps}.

\edits{We put our results in the context of other attempts to map the completeness of \gaia DR2 in Sec. \ref{sec:discussion}, and also discuss how the methodology presented in this paper could be extended to the parallax and proper motion, colour photometry, variable star and radial velocity subsets of \gaia DR2. We end by presenting our new \textsc{Python} package \textsc{selectionfunctions} which will allow the reader to easily incorporate our selection functions in their own work.}

\section{Methodology}
\label{sec:methodology}

\begin{figure*}
	\centering
	\includegraphics[width=1.\linewidth,trim=0 0 0 0, clip]{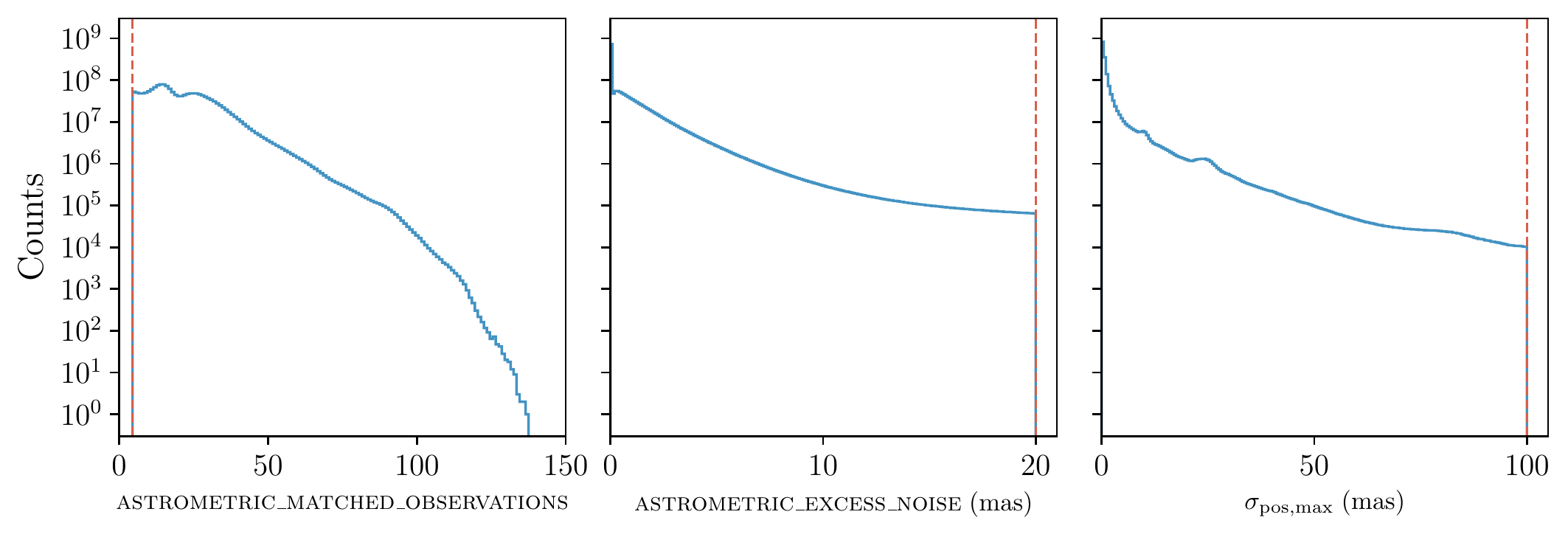}
	\caption{The distribution of all sources in \gaia DR2 (blue) with the cuts (red, dashed) used to select which sources were included in \gaia DR2 \citep{Lindegren2018}.}
	\label{fig:cutjustification}
\end{figure*}

This work tackles the selection function of the \gaia DR2 source catalogue, which is the catalogue of 1,692,919,135 sources detected by \gaia that satisfied the basic astrometric quality cuts. Sources were included if their five parameter astrometric solution (position, parallax and proper motion) satisfied the quality cuts given by Eq.~11 of \citet{Lindegren2018} or -- failing that -- if their two parameter astrometric solution (position only) satisfied the quality cuts
\begin{equation}
\begin{cases}
      \textsc{astrometric\_matched\_observations} &\geq 5, \\
      \textsc{astrometric\_excess\_noise} &< 20\;\mathrm{mas}, \\
      \sigma_{\mathrm{pos,max}} &< 100\;\mathrm{mas},
\end{cases}
\label{eq:lindegrentwo}
\end{equation}
where \textsc{astrometric\_matched\_observations} is the number of field-of-view transits that contributed measurements to the astrometric solution, \textsc{astrometric\_excess\_noise} quantifies the goodness-of-fit of the solution, and $\sigma_{\mathrm{pos,max}}$ is the semi-major axis of the position uncertainty ellipse \citep{Lindegren2018}. In practise, if a source has a valid five parameter astrometric solution then the two parameter astrometric solution would also have been valid, and so we can consider the cuts in Eq. \ref{eq:lindegrentwo} to define the selection of sources for the \gaia DR2 source catalogue. \edits{We show in Fig. \ref{fig:cutjustification} histograms of \textsc{astrometric\_matched\_observations}, \textsc{astrometric\_excess\_noise} and $\sigma_{\mathrm{pos,max}}$ for all sources in \gaia DR2.} While the cuts in the latter two of these occur far out in the tail of the distribution, the cut in \textsc{astrometric\_matched\_observations} occurs at the peak of the distribution. \citet{Lindegren2018} notes that most of the sources excluded by the cuts in Eq. \ref{eq:lindegrentwo} are spurious, and we therefore conclude that for genuine sources the only effective cut is $\textsc{astrometric\_matched\_observations}\geq5$. \edits{There is one additional requirement\footnote{\url{https://gea.esac.esa.int/archive/documentation/GDR2/Catalogue_consolidation/chap_cu9cva/sec_cu9cva_consolidation/ssec_cu9cva_consolidation_ingestion.html}} that there must have been at least ten $\textsc{phot\_g\_n\_obs}$ ($G$ CCD transits), however this cut is much weaker than that on \textsc{astrometric\_matched\_observations} because each focal plane transit can result in as many as nine $\textsc{phot\_g\_n\_obs}$.}

The approach we have developed to calculate the \gaia DR2 selection function is entirely novel, and so to aid clarity we give brief definitions of the key terminology here. A \textit{source} is any object which \gaia would report as a single entry in the DR2 catalogue (used interchangeably with \textit{star}). An \textit{observation} of a source is a single occasion on which a source could be seen by \gaia, because it is within \gaia's field-of-view. A \textit{detection} is an observation during which \gaia notices the presence of the source and obtains a measurement that subsequently contributes to the \edits{astrometric solution in} the \gaia DR2 catalogue. \edits{Observations only result in detections if the source is acquired by the SkyMapper CCD and selected to have a window tracked on-board, the source is confirmed by the AF1 CCD, the telemetry is successfully sent to the ground and processed by DPAC, and the measurements pass all the minimum astrometric and photometric quality criteria to make it to the Gaia DR2 catalogue.} The number of detections is given in \gaia DR2 as \textsc{astrometric\_matched\_observations} and is less than or equal to the number of observations. The \textit{detection probability} is the probability that \gaia will detect a source on a given observation. The \textit{selection function} is the probability that a source is detected on at least five observations of that source. The core assumption of this work is that a source having been detected at least five times is the sole requirement for inclusion in \gaia DR2.

\subsection{Flipping biased coins}
\label{sec:bias}
Our methodology is rooted in a simple problem that is often used to introduce Bayesian statistics: how do you determine the bias of a weighted coin? Suppose we flip a coin $n$ times and observe $k$ heads and $n-k$ tails. If the coin is fair, then the probability of observing $k$ heads is given by
\begin{equation}
\operatorname{P}(k|n) = \binom{n}{k}\left(\frac{1}{2}\right)^{n}, \label{eq:faircoin}
\end{equation}
because there are $\binom{n}{k}$ ways to pick which coin flips resulted in heads and on each flip there is a $\frac{1}{2}$ chance of either heads or tails. If the coin is biased, such that the probability of heads on any one flip is given by $\theta$, then the probability of $k$ heads is given by
\begin{equation}
\operatorname{P}(k|n,\theta) = \binom{n}{k}\theta^k(1-\theta)^{n-k}, \label{eq:binomial}
\end{equation}
which simplifies to Eq. \ref{eq:faircoin} in the case $\theta=\frac{1}{2}$. This is the probability mass function of a Binomial random variable.

If we do not know the precise bias of the coin, then we will want to quantify our knowledge in terms of a prior $\operatorname{P}(\theta)$, which gives our a priori belief that $\theta$ takes each possible value between 0 and 1. A commonly-used choice is the Beta distribution
\begin{equation}
    \operatorname{P}(\theta|\alpha,\beta) = \frac{\theta^{\alpha-1}(1-\theta)^{\beta-1}}{\operatorname{B}(\alpha,\beta)},
\end{equation}
where the denominator is the beta function\footnote{The Beta distribution, the beta function and the $\beta$ variable are all separate entities.}
\begin{equation}
    \operatorname{B}(\alpha,\beta) = \int_0^1 x^{\alpha-1}(1-x)^{\beta-1}\mathrm{d}x=\frac{\Gamma(\alpha)\Gamma(\beta)}{\Gamma(\alpha+\beta)}.
\end{equation}
The Beta distribution is the conjugate prior\footnote{A distribution is a conjugate prior of a likelihood function if the posterior distribution belongs to the same probability distribution family as the prior distribution.} of the Binomial distribution, meaning that if your prior belief is that $\theta\sim\operatorname{Beta}(\alpha,\beta)$ and you subsequently observe $k$ heads from $n$ flips, then your posterior belief should be that $\theta\sim\operatorname{Beta}(\alpha+k,\beta+n-k)$. This pairing is known as the Beta-Binomial model. The probability of observing $k$ heads from $n$ flips is then, after marginalising over $\theta$,
\begin{equation}
\operatorname{P}(k|n,\alpha,\beta)=\binom{n}{k}\frac{\operatorname{B}(\alpha+k,\beta+n-k)}{\operatorname{B}(\alpha,\beta)}.
\end{equation}
If $\alpha=1$ and $\beta=1$ then the Beta prior is equivalent to the Uniform distribution over $(0,1)$, which corresponds to having no knowledge about the bias of the coin.

In Appendix \ref{sec:betabinomial} we have illustrated the range of possible shapes of the Binomial and Beta distributions, which may be of interest to readers unfamiliar with these distributions.

If we had multiple coins $i$, then we could either assume that they all have the same bias $\theta\sim\operatorname{Beta}(\alpha,\beta)$, or that they each have their own bias $\theta_i\sim\operatorname{Beta}(\alpha,\beta)$. In the former case the Beta distribution quantifies our uncertainty on the true bias of the coins, while in the latter case it quantifies the intrinsic spread in the biases.

The connection between this problem and the problem of the \gaia selection function is immediate: we are attempting to quantify the probability $\theta$ that \gaia detects a star $k$ times out of $n$ observations, which is the same statistical problem as quantifying the probability $\theta$ that a coin comes up heads $k$ times out of $n$ flips. However, \gaia DR2 only reports sources if $k\geq5$, which is analogous to flipping a number of coins multiple times but discarding any data from the coins where the number of heads was fewer than five. An additional complication is that the detection probability is likely to depend on the properties of the star, which requires us to stretch our analogy to breaking point in the form of the following fictitious story.


Suppose that we suspect that the Royal Mint has been producing biased pennies for the last \edits{fifty} years. Even worse, the bias of every coin is different, and the distribution of those biases appears to be changing from year to year. To investigate this troubling phenomenon, we establish the Penny Processing and Analysis Consortium and ask volunteers all over Britain to gather their pennies, flip them until they get bored, and send us a postcard with one line for each coin giving the number of flips, the number of heads and the year of minting. After the British public has dutifully flipped 1,692,919,135 pennies\footnote{A total of 22,820,373,346 pennies had been minted up until the 14$^{\mathrm{th}}$ October 2019 (\url{https://www.royalmint.com/currency/uk-currency/mintages/1-penny/}).} a reported total of 48,421,126,164 times, we are dismayed to discover that none of the coins in our dataset had come up heads less than five times. This seems unlikely to be true, given that 52,871,273 of the coins had shown heads exactly five times. It emerges that one of our researchers had carelessly stated during a TV interview that `we expect very few coins to have come up heads less than five times', and the public had taken this to heart, discarding the data about a coin if there were fewer than five heads. It now falls to us to infer the changing distribution of the penny biases from year-to-year, accounting for the missing data.

This story describes the exact same statistical problem that we have solved in this work. In place of the year of minting we binned the stars by their brightness (and later their neighbouring stellar density), and computed the distribution of detection probabilities in each of these bins, subject to the data truncation that none of the reported stars had fewer than five detections.

\subsection{Statistical framework}
\label{sec:framework}
Suppose that for each of the 1,692,919,135 sources in \gaia DR2 (labelled $i$) we have the number of detections $k_i$, the number of times that that source crossed the field-of-view $n_i$ and the $G_i$ magnitude. At the time of each observation there will be some probability that \gaia will successfully detect the star. Our first simplifying assumption is that the probability of detection is the same at every observation and we term this probability $\theta_i$. It is possible that the probability could vary with time (for instance, the amount of scattered light from the frayed edges of the solar shield will be a function of the phase of \gaia's orbit) and thus our $\theta_i$ will be an approximate average of the true probabilities. We discuss the difficulties of weakening this assumption in Sec. \ref{sec:poissonbinomial}.

We assumed that the only selection for a star to be included in \gaia DR2 was that $k_i\geq5$. If that selection had not been applied, then we could have modelled $k_i$ as a Binomial random variable with $n_i$ observations and probability of success $\theta_i$, and thus the probability mass function would be
\begin{equation}
P(k_i|n_i,\theta_i) = \begin{cases} 
\binom{n_i}{k_i} \theta_i^{k_i} (1-\theta_i)^{n_i-k_i} &\mathrm{for}\;k_i \geq 0, \\ 
0 &\mathrm{otherwise}.\end{cases}
\end{equation}
The selection $k_i\geq5$ acts to adjust the probability mass function:
\begin{equation}
    P(k_i|n_i,\theta_i,S) = \begin{cases} 
    \frac{1}{P(k\geq5|n_i,\theta_i)}\binom{n_i}{k_i} \theta_i^{k_i} (1-\theta_i)^{n_i-k_i} &\mathrm{for}\;k_i \geq 5, \\ 
    0 &\mathrm{otherwise},\end{cases} \label{eq:pmf}
\end{equation}
where $P(k\geq5|n_i,\theta_i) = \operatorname{I}_{\theta_i}(5,n_i-4)$ is an \edits{incomplete beta function},
\begin{equation}
    \operatorname{I}_{\theta}(\alpha,\beta) = \int_0^{\theta} x^{\alpha-1}(1-x)^{\beta-1}\mathrm{d}x,
\end{equation}
and is the survival function of the Binomial distribution.

We further assumed that the probability $\theta_i$ will depend primarily on the brightness of the star $G$ (although in Sec. \ref{sec:crowding} we additionally consider the neighbouring source density) \edits{and considered two different models of that dependency}. Our simple model (hereafter Model T) was to assume $\theta_i$ is directly a function of $G_i$, and thus that all stars with the same magnitude $G$ have the same detection probability $\theta = T(G)$. Our more realistic model (hereafter Model AB) was to assume that the $\theta$ for each star was drawn from a Beta distribution with parameters $A(G)$ and $B(G)$ that are each functions of the brightness. Model AB can account for stars of the same magnitude having a distribution of detection probabilities \edits{and allows that distribution to change with magnitude.}

\begin{figure}
	\centering
    \includegraphics[width=1.\linewidth,trim=0 10 0 10, clip]{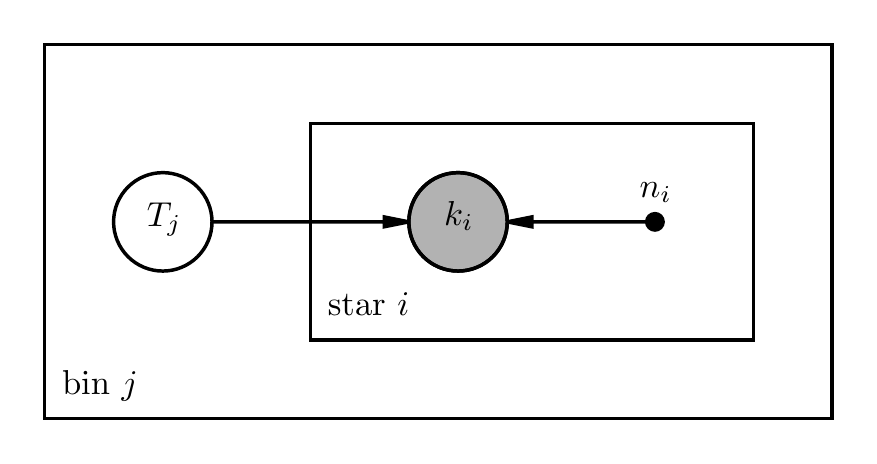}
    (a) Model T
    \includegraphics[width=1.\linewidth,trim=0 10 0 0, clip]{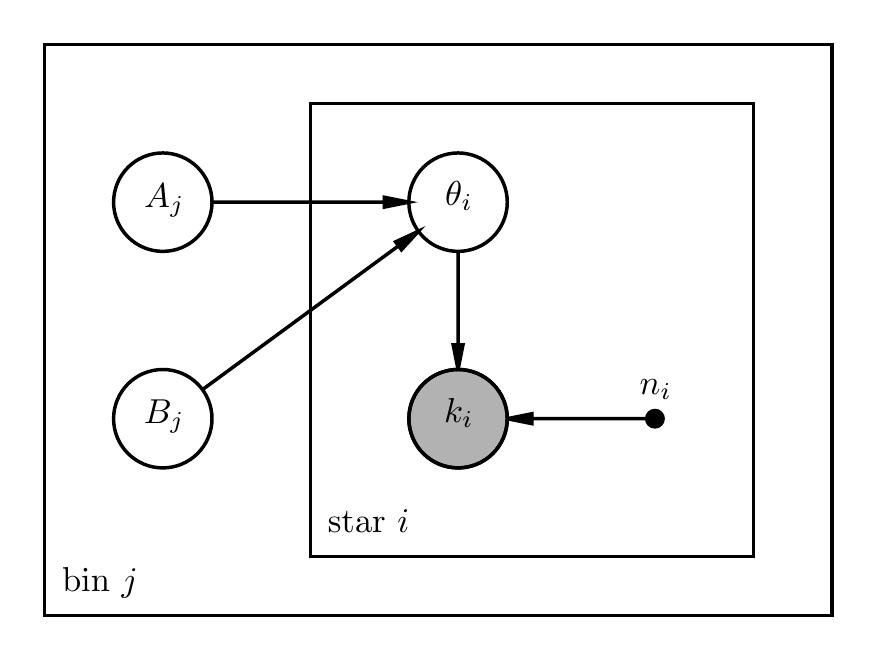}
    (b) Model AB
	\caption{Plate diagrams for our two models. The stars in \gaia DR2 are divided into bins by their $G$ magnitude and their number of detections $k$ is assumed to be Binomial-distributed with number of observations $n$ and probability of detection $\theta$. \textbf{Top:} In Model T the probability of detection is assumed to be the same $T_j$ for all the stars in each bin. \textbf{Bottom:} In Model AB the the probability of detection of each star in the bin is assumed to have been drawn from a Beta distribution with parameters $A_j$ and $B_j$.}
	\label{fig:plate}
\end{figure}

We opted to model the functions $T(G)$, $A(G)$ and $B(G)$ as piecewise-constant over decimag bins in $G$ from $1.7$ to  $23.5$, where the value of each function in each of the 218 bins $j$ is given by the hyper-parameters $T_j$, $A_j$ and $B_j$. We additionally require that $0<T(G)<1$, $A(G)>0$ and $B(G)>0$ due to the constraints on the parameters of the Binomial and Beta distributions. In practice, we further restricted the domains to $10^{-1}<A(G),B(G)<10^{4}$ to allow us to pre-compute the numerically expensive incomplete beta function appearing in Eq. \ref{eq:pmf}, but note that our posteriors only run up against these boundaries in bins with very few stars. We have illustrated the relationship between these parameters and the observables through the plate diagrams in Fig. \ref{fig:plate}.

\subsection{Computing the number of observations}
\label{sec:n}

\begin{figure*}
	\centering
	\includegraphics[width=1.\linewidth,trim=70 110 35 110, clip]{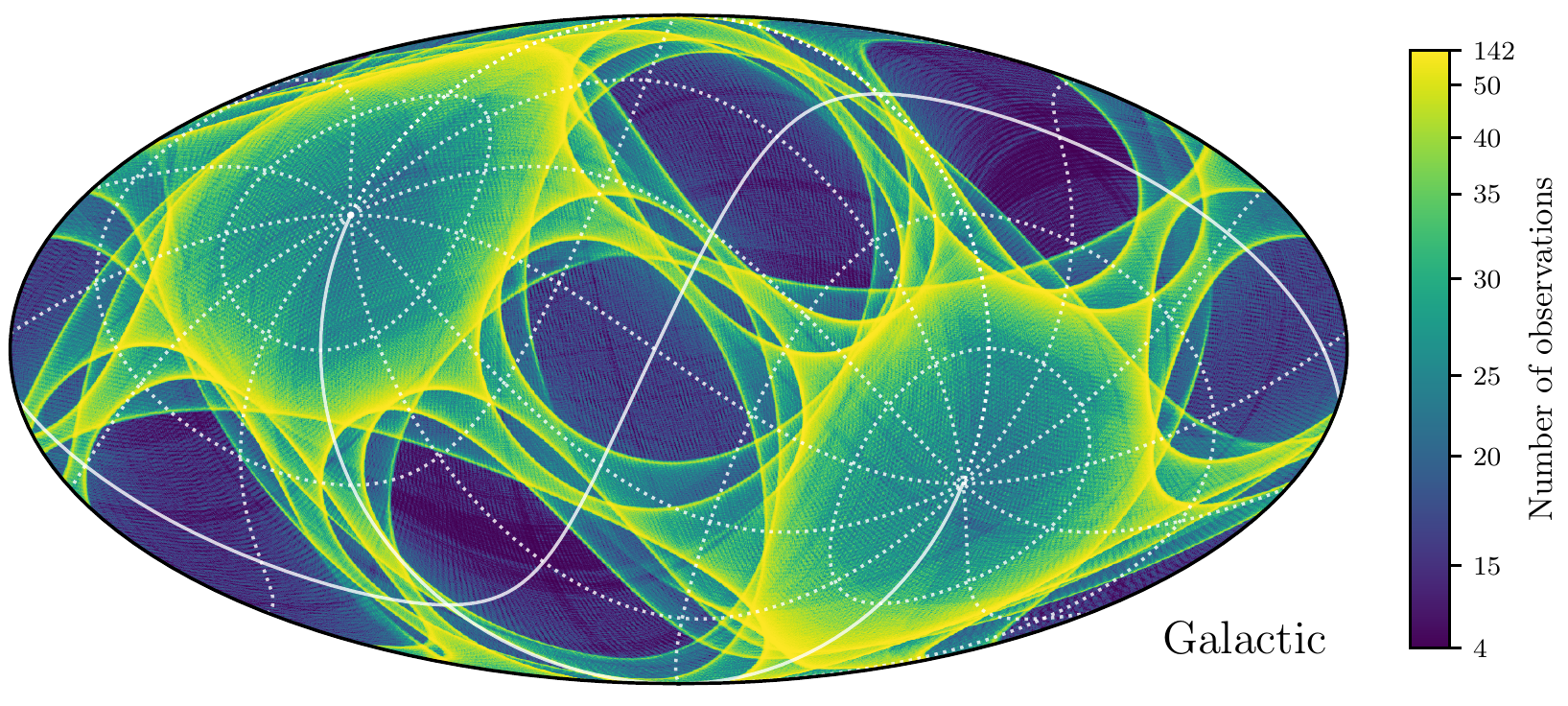}
	\caption{A \edits{Galactic} HEALPix map of the estimated number of times that \gaia looked at each location on the sky. The graticule marks out lines of longitude and latitude in the Ecliptic coordinate frame, which is the native frame of the \gaia scanning law. The colourmap has been histogram-equalised to increase the contrast. This figure is similar to Fig. 2 of \citet{Boubert2020}, but \edits{with the observations during periods that were excluded from the astrometric data-taking removed.}}
	\label{fig:map}
\end{figure*}

Our methodology requires that we know the number of times $n_i$ that each of the stars in \gaia DR2 was observed, i.e. transited across either the preceding or following \gaia field-of-view (FoV). \edits{In Paper I of this series \citep{Boubert2020}, we precisely determined the \gaia scanning law over the period of DR2 using the recently published DR2 nominal scanning law\footnote{\url{https://www.cosmos.esa.int/web/gaia/scanning-law-pointings}} together with the detection times of each of the 550,737 variable sources with DR2 light-curves.} There are stretches of time during the 22 months covered by DR2 when \gaia was not able to obtain astrometric measurements or the astrometric measurements that were taken were erroneous, for example, due to the Ecliptic Pole Scanning Law, decontamination procedures, micro-meteroid impacts or station-keeping maneuvers \citep{Lindegren2018}, and it is vital to account for these gaps. \edits{We use the astrometric data gaps recently provided by the \gaia DPAC\footnote{\url{https://www.cosmos.esa.int/web/gaia/dr2-data-gaps}} and refer the interested reader to Paper I for more details.} We use the same methodology as detailed in Paper I to predict $n_i$ for all of the individual sources in \gaia DR2.

\edits{A crucial requirement of our methodology is that the $G$ magnitude reported in the \gaia catalogue is accurate, because the proportion of observations that result in detections will be much lower for fainter sources. If a truly bright source was reported to be faint, then the higher detection efficiency could greatly bias our posterior on the selection function. In Paper I we identified that almost all of the faintest stars in \gaia DR2 have spurious magnitudes resulting from a miscalibration due to a thunderstorm over Madrid, with most of the remainder being attributed to a similar second miscalibration event. We therefore do not consider the 72,785,162 sources with a predicted observation in the time periods $\mathrm{OBMT}=1388{-}1392\;\mathrm{rev}$ and $\mathrm{OBMT}=2211{-}2215\;\mathrm{rev}$ in the remainder of this work.} 

When using the results of this work to map the selection function on the sky, we assume that the number of observations of a source is approximately equal to the number of observations of the centre of the nearest pixel of an \edits{\textsc{nside}=4096}, Equatorial, nested HEALPix map, which we show in Fig. \ref{fig:map}. This map reveals the intricate overlaps between the successive precessing scans of \gaia \edits{, which we will show in later sections result in equally intricate variations in the completeness of \gaia DR2 across the sky.}

\subsection{Implementation of Bayesian model}

In this section we detail the implementation of the models described in Sec. \ref{sec:framework}. We first binned the sources in \gaia DR2 by their $G$ magnitude, by their number of detections $k$ and by the number of observations $n$ we predicted in the previous section. Paradoxically, there were \edits{619,272} sources where their predicted $n$ was less than $k$\edits{, likely due to a combination of small uncertainties in our prediction of the numbers of observation and spurious duplicate detections of the bright stars (as discussed in Paper I). Rather than discard these sources, in Appendix \ref{sec:deconvolution} we deconvolve the distribution of all sources in $(n,k)$ for each slice in $G$ to account for these processes. We have verified that the results in the remainder of this paper are not substantially changed if we instead discard the sources with $n<k$ and leave the distribution of the other stars unchanged.} We note that the models shown in Fig. \ref{fig:plate} are entirely independent between the magnitude bins, and so we could calculate the posterior model in each bin separately. Within each magnitude bin the likelihood is only a function of $n$, $k$ and the hyper-parameters of the model, and thus we can calculate the likelihood once for each combination of $n$ and $k$ and then multiply the log-likelihood by the number of sources with that $n$ and $k$. For each of the models and each of the bins we use the affine invariant Markov chain Monte Carlo (MCMC) ensemble sampler \citep{Goodman2010} implemented in \textsc{emcee} \citep{Foreman2013,Foreman2019} to draw samples from the posterior over the hyper-parameters. In each case we used 32 walkers, drew samples until the chain was at least 50 times longer than the chain's autocorrelation time $\tau$, and discarded the first $5\lceil\tau\rceil$ samples as `burn-in'.

\subsubsection{Likelihood and priors for Model T}
The likelihood under Model T of $k$ \edits{reported detections} for a source in magnitude bin $j$ with $n$ \edits{predicted observations} is
\begin{equation}
    P(k|n,T_j) = \begin{cases} 
    \frac{1}{\operatorname{I}_{T_j}(5,n-4)}\binom{n}{k} T_j^k (1-T_j)^{n-k} &\mathrm{for}\;k \geq 5, \\ 
    0 &\mathrm{otherwise}.\end{cases} \label{eq:modelT}
\end{equation}
We assumed a uniform prior over $(0,1)$ for $T_j$ in each magnitude bin. The incomplete beta function is computationally-expensive to evaluate and hence we opted to pre-compute the value of $\operatorname{I}_x(5,n-4)$ at 10,000 equally spaced points $x\in(0,1)$ for each possible $n$. We then interpolated this grid with a cubic spline in place of explicitly evaluating the incomplete beta at each MCMC step.

If the selection $k\geq5$ is not applied then Model T would be identical to the Beta-Binomial model described in Sec. \ref{sec:bias}. Suppose we have a biased coin with a prior on the probability of success $p \sim \operatorname{Beta}(\alpha,\beta)$, where $\alpha$ and $\beta$ are fixed. We then conduct $N$ trials and in each trial $i$ conduct $n_i$ flips and observe $k_i$ successes with $k_i\sim\operatorname{Bin}(n_i,p)$. The posterior is simply $p\sim\operatorname{Beta}(\alpha+\Sigma k_i,\beta + \Sigma [n_i-k_i])$. We expect that \gaia will detect sources of intermediate brightness on almost every transit, and thus over this range in magnitude the Beta-Binomial model will give a close approximation to Model T because the incomplete beta function will be close to 1.

\subsubsection{Likelihood and priors for Model AB}
The Model AB is hierarchical as there is an additional parameter $\theta$ for each source that is drawn from a Beta distribution with hyper-parameters $A_j$ and $B_j$:
\begin{equation}
    P(\theta|A_j,B_j) = \frac{\theta^{A_j-1}(1-\theta)^{B_j-1}}{\operatorname{B}(A_j,B_j)}.
\end{equation}
The likelihood under Model AB of $k$ detections for a source in magnitude bin $j$ with $n$ transits is then
\begin{equation}
    P(k|n,\theta) = \begin{cases} 
    \frac{1}{\operatorname{I}_{\theta}(5,n-4)}\binom{n}{k} \theta^k (1-\theta)^{n-k} &\mathrm{for}\;k \geq 5, \\ 
    0 &\mathrm{otherwise}.\end{cases}
\end{equation}
Obtaining the posterior of a Bayesian model with tens of millions of parameters is not \edits{generally} feasible and so we were forced to marginalise over the $\theta$ parameter of each source. For each possible combination of $n$ and $k$ we numerically evaluated the integral
\begin{equation}
    J_{n,k}(A,B) = \int_0^1 \frac{\theta^{A+n-1}(1-\theta)^{B+n-k-1}}{\operatorname{I}_{\theta}(5,n-4)}\mathrm{d}\theta
\end{equation}
over a grid in $(A,B)$ with logarithmic spacings between $10^{-1}$ and $10^{4}$ in each direction. We then implemented a bivariate cubic spline in $(\log_{10}A,\log_{10}B)$ for each $n$ and $k$. The likelihood is thus simplified to
\begin{equation}
    P(k|n,A_j,B_j) = \begin{cases} 
    \binom{n}{k}\frac{J_{n,k}(A_j,B_j)}{\operatorname{B}(A_j,B_j)} &\mathrm{for}\;k \geq 5, \\ 
    0 &\mathrm{otherwise}.\end{cases} \label{eq:modelAB}
\end{equation}
We adopted \edits{log-}uniform priors for $A_j$ and $B_j$ over $(10^{-1},10^{4})$ to ensure that the MCMC walkers do not explore outside our pre-computed grid. The limits of this grid were determined through experimentation and we can confirm that the posteriors on $A_j$ and $B_j$ only extend to the limits in the case of bins with small numbers of sources (and thus broad unconstrained posteriors).

If the selection $k\geq5$ is not applied then the integral in the likelihood has the closed form $J_{n,k}(A,B)=\operatorname{B}(A+n,B+n-k)$. Unlike in the Beta-Binomial case discussed in the previous section this does not lead to an analytic posterior, but this simpler model does provide a valuable sanity check on Model AB.

\section{Results}
\label{sec:results}

\subsection{Posterior selection functions}
\begin{figure}
	\centering
    \includegraphics[width=1.\linewidth,trim=0 0 0 0, clip]{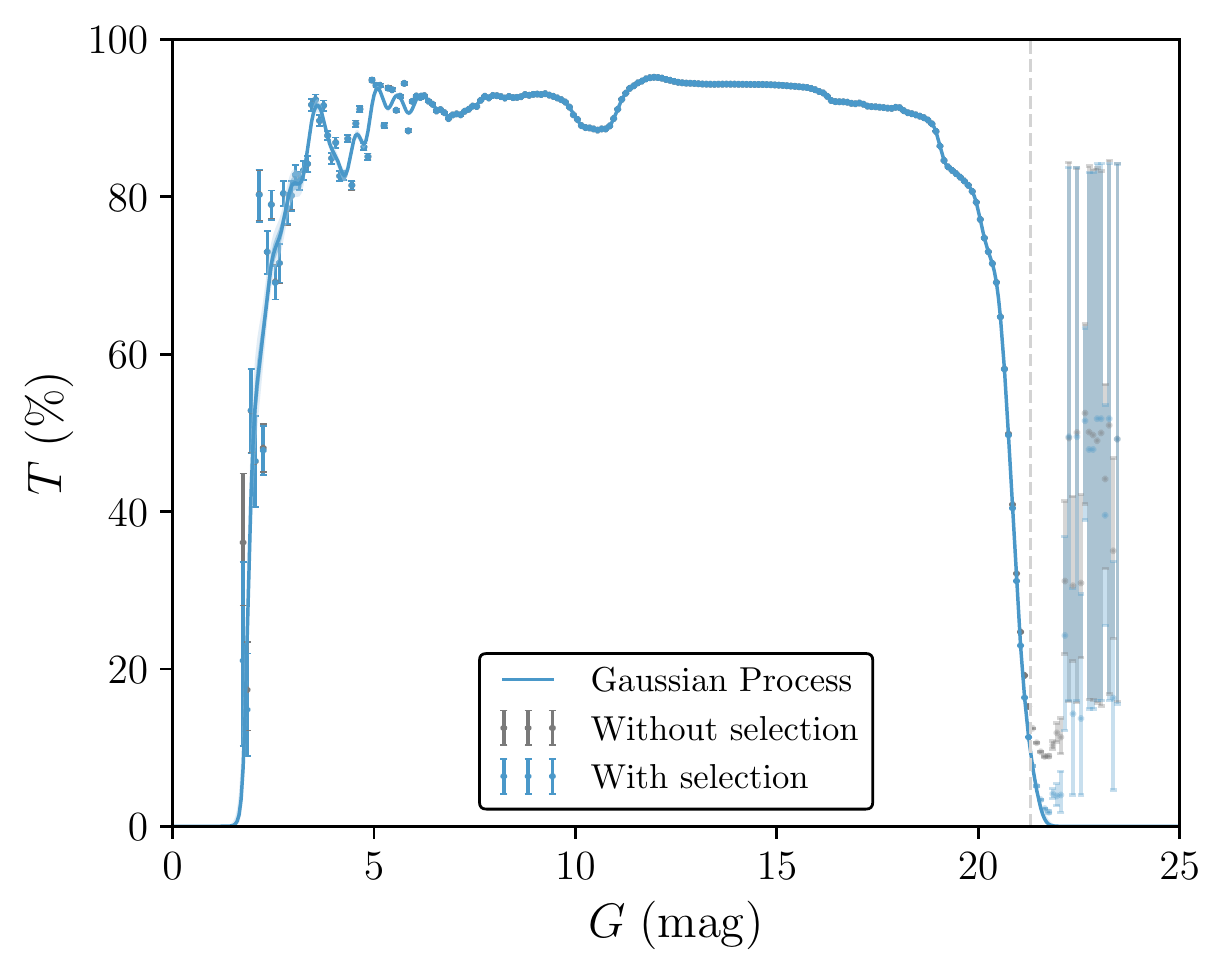}
    (a) Posterior on Model T
    \includegraphics[width=1.\linewidth,trim=0 0 0 0, clip]{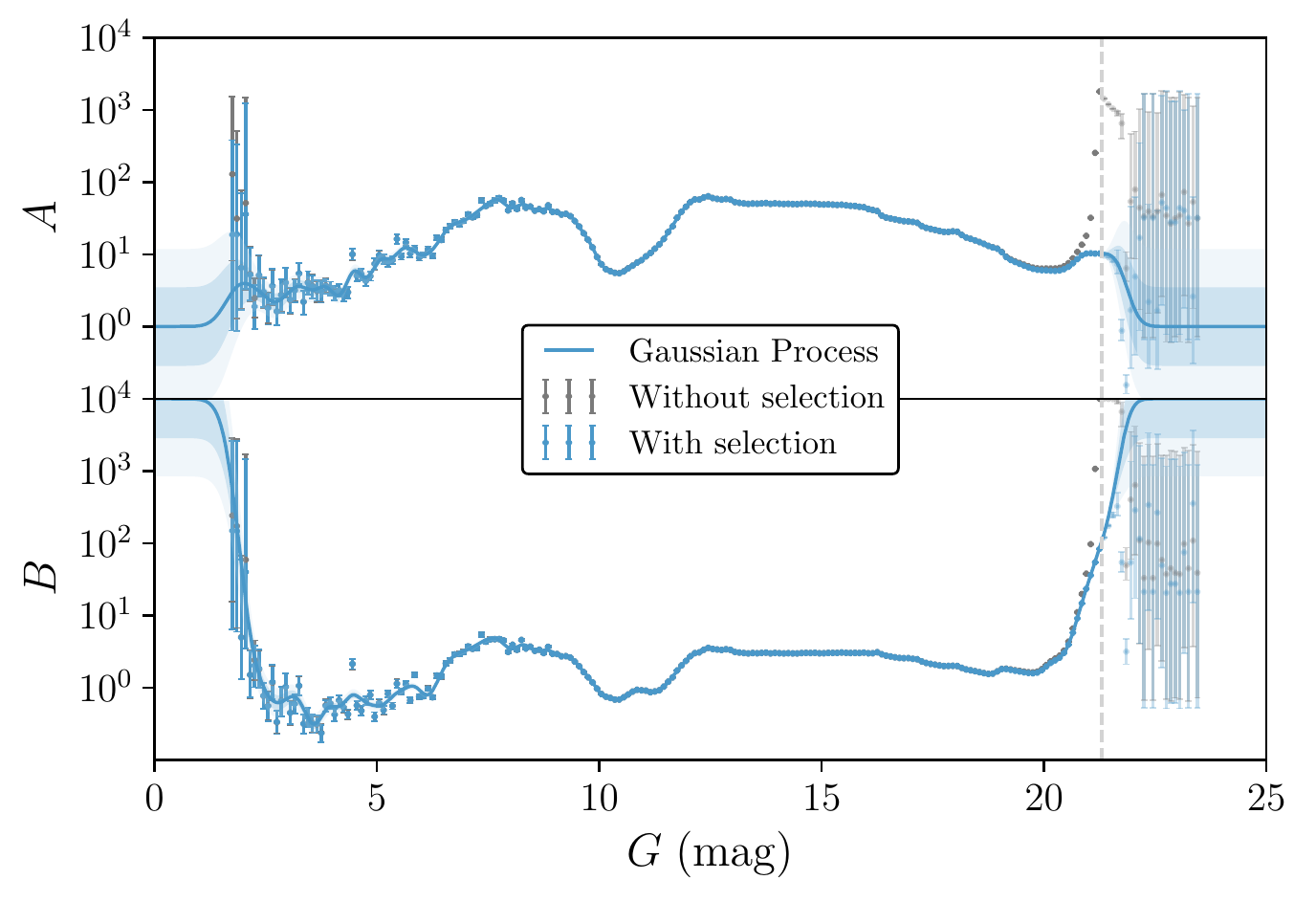}
    (b) Posterior on Model AB
	\caption{The posteriors on our two models of the probability that \gaia detects a source each time the source crosses the field-of-view. In both panels we show the \edits{median value with the $1\sigma$ error-bars. We also show our post-hoc Gaussian Process fits which continuously interpolate these data points and allow us to extrapolate at the bright and faint end. The faint data points to the right of the dashed line at $G=21.3$ are those which are possibly biased due to sources with miscalibrated magnitudes and which we discarded before fitting the Gaussian Processes.} \textbf{Top:} In Model T, the detection probability is assumed to be a deterministic function of the magnitude of the source. \textbf{Bottom:} In Model AB, the detection probability of a source by \gaia is drawn from a Beta distribution whose parameters are deterministic functions of the magnitude of the star. To allow for comparison with Model T, we show the resulting posterior on the detection probability in the top panel of Fig. \ref{fig:completeness}.}
	\label{fig:modelposterior}
\end{figure}

We computed the posteriors of both models with and without the $k\geq5$ selection, and show the median with \edits{$1\sigma$ error-bars} on $T$, $A$ and $B$ as a function of magnitude in Fig. \ref{fig:modelposterior}. The \edits{uncertainties} are large at both the bright ($G<8$) and faint ($G>22$) ends due to the small number of \gaia DR2 sources per decimag bin at these magnitudes. The difference between the posteriors with and without the $k\geq5$ selection is only apparent when the posterior indicates that the typical detection probability is less than 20\% ($T<0.2$ or $A/(A+B)<0.2$), because the odds of a star having $k<5$ are otherwise vanishingly small, given that that the average number of observations is \edits{28}.

\edits{As mentioned in Sec. \ref{sec:n}, we have not included any of the sources observed during the miscalibrated periods identified in Paper I and thus have removed almost all of the sources fainter than $G=22.1$, causing the large uncertainty in all three parameters at the faint end. There are likely to be some remaining miscalibrated sources which may bias our results, given that there are still sources reported to be as faint as $G=23.1$ in our sample. We note, for example, that $T(G)$ is smoothly declining over $20<G<21.7$, but then begins to rise again. We opt to exclude all datapoints for magnitudes fainter than $G=21.3$ from our analysis, motivated by empirical investigations of the run of each parameter with $G$ magnitude.}

\edits{We chose to model $T$, $A$ and $B$ as piece-wise to expedite the Bayesian inference by making it independent between each magnitude bin, but this is of course only an approximation. It would have been preferable to model the continuous run of $T$, $A$ and $B$ with $G$ across all magnitudes simultaneously. This would also result in a more precise estimation of $T$, $A$ and $B$, because their value at a particular $G$ would be informed by their value at neighbouring magnitudes. We opted to model the data-points shown in Fig. \ref{fig:modelposterior} as independent Gaussian Processes with a squared-exponential kernel. This required us to transform the $T$ data-points from $(0,1)$ to $(-\infty,+\infty)$ through a logit transform and the $A$ and $B$ data-points from $(0,+\infty)$ to $(-\infty,+\infty)$ through a log$_{10}$ transform, with the uncertainties being propagated by transforming the 16\% and 84\% percentiles and then averaging their distance from the transformed magnitudes. Through experimentation we fixed the length-scale of the Gaussian Processes at $0.3\;\mathrm{mag}$ and fixed the variances at $1.0$ for $T$ and $0.3$ for $A$ and $B$ (though note that these variances are in the logit and log$_{10}$ transformed spaces). An advantage of using Gaussian Processes is that we can use them to extrapolate at the bright and faint end. We chose the means of the Gaussian Processes to be -10, 0 and 4 for $T$, $A$ and $B$ respectively in order to ensure that the posteriors on the detection probability tended towards zero away from regions constrained by data. We evaluated the conditional distribution of these Gaussian Processes at 501 points uniformly spaced by $0.05\;\mathrm{mag}$ over $0\leq G \leq25$ and applied the reverse of the logit and log$_{10}$ transforms to obtain the median and $1$ and $2\sigma$ regions shown in Fig. \ref{fig:modelposterior}. We use interpolation of these points to obtain the results in the remainder of this work.}

The top panel of Fig. \ref{fig:completeness} shows the posterior distribution of the detection probability $\theta$ under Model AB\edits{, where we have taken the mean conditional value of the Gaussian Processes over $A(G)$ and $B(G)$ at each magnitude and computed the percentiles of the detection probability from the corresponding $\operatorname{Beta}(A,B)$ distribution. The contours shown in Fig. \ref{fig:completeness} thus only illustrate the spread of detection probabilities under our most likely model at each magnitude, and do not incorporate any uncertainty in the model parameters.}

\subsection{Completeness as a function of magnitude}
\label{sec:completeness}

\begin{figure*}
	\centering
	\includegraphics[width=1.\linewidth,trim=0 0 0 0, clip]{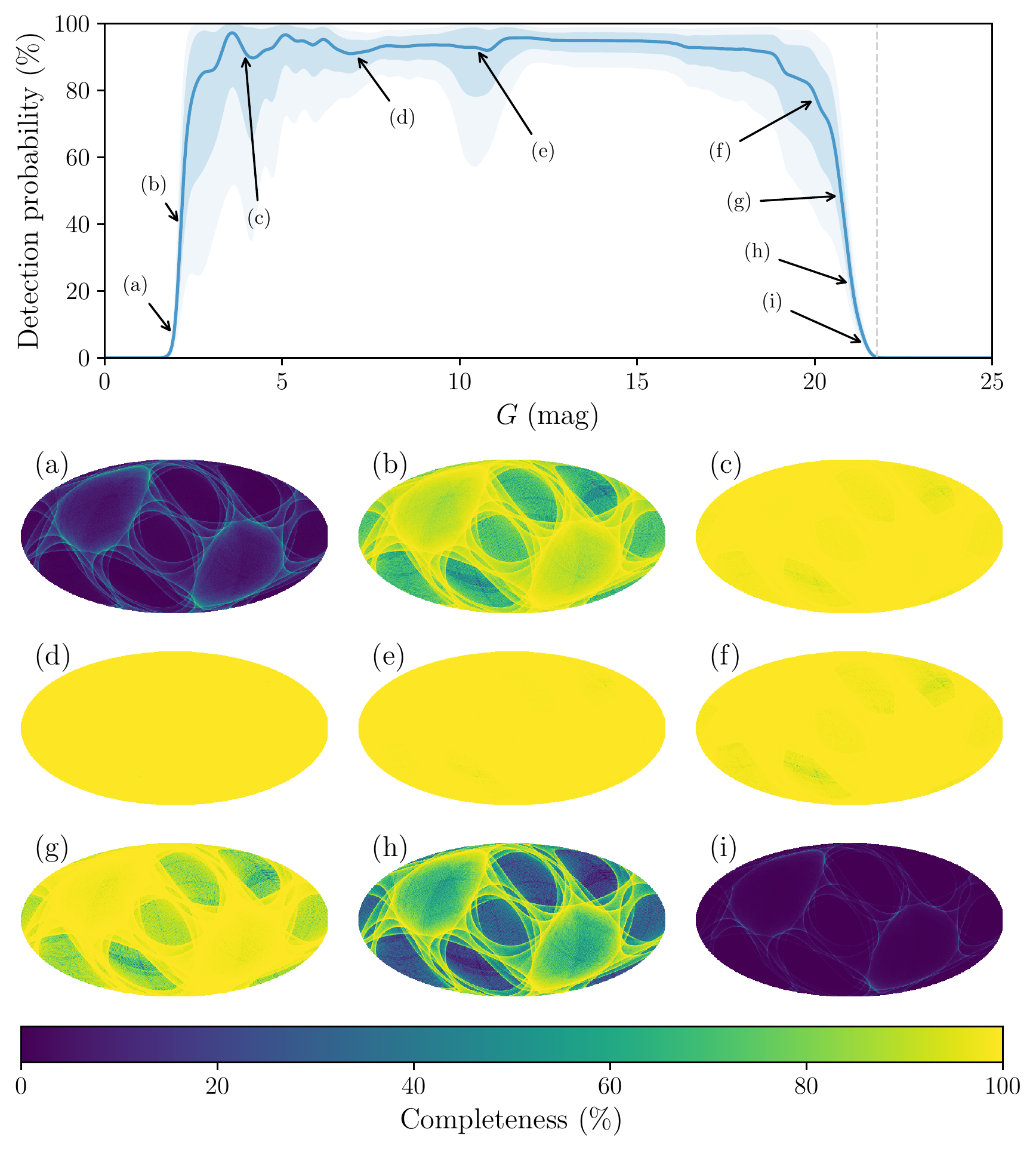}
	\caption{Our Model AB predicts the distribution of probabilities that a source is detected by \gaia as a function of magnitude, which we illustrate in the top panel with the median and $1$ and $2\sigma$ regions. These distributions can be used to compute the completeness of \gaia DR2 with respect to sources which are of magnitude $G$ and which have received $n$ observations. In the bottom panel, we show a grid of \edits{Galactic} completeness maps at the magnitudes labelled in the top panel. An \href{https://www.gaiaverse.space/publications/paper-ii}{interactive version} of this figure is available online (please note that the static HTML behind the webpage will take up to thirty seconds to download and decompress). \edits{We stress that the results shown here do not account for the deleterious effects of crowding on \gaia's completeness (as considered further in Sec. \ref{sec:crowding}) and so should be treated as a biased indication of how the completeness changes with source magnitude and the number of observations.}}
	\label{fig:completeness}
\end{figure*}

Our inferred detection probabilities can be used to predict the completeness of \gaia DR2 as a function of magnitude and number of observations. The completeness is simply the fraction of stars with magnitude $G$ that were observed $n$ times and were detected on at least five occasions. We focused our attention on the more \edits{general} Model AB, where the probability of detection $\theta$ of each source is assumed to be Beta-distributed with parameters $A(G)$ and $B(G)$ which are each functions of magnitude. The completeness \edits{increases} with the number of observations because more observations equates to more occasions on which a source can be detected. We demonstrated in Fig. \ref{fig:map} that the number of observations that a source receives strongly varies with position on the sky, and thus the completeness of \gaia DR2 at a given magnitude can change from 0\% to 100\% depending \edits{upon} where the source is on the sky. Predicting the completeness from Model AB is not trivial and we outline the procedure we followed in Appendix \ref{sec:completenessAB}.

In the bottom panels of Fig. \ref{fig:completeness}, we show maps of the fraction of stars that would have been detected by \gaia, at \edits{selected} magnitudes where the distribution of detection probabilities appears to be changing. As expected, the high detection probability over $3<G<20$ implies that \gaia is essentially complete over the entire sky at these magnitudes. The dip between $10<G<12$ has a negligible effect on the completeness. The completeness does drop at the extreme bright end $G<3$ and at the faint end, falling from $100\%$ to $0\%$ over $20.0<G<21.5$. 

The other features in the top panel of Fig. \ref{fig:completeness} can be explained by \edits{technical details of the \gaia instrumentation}. The decline at magnitudes brighter than (d) is due to saturation for $G<6$. \edits{The treatment of bright stars (particularly of very bright stars $G<3$) was preliminary in \gaia DR2 and will be improved in future data releases \citep{Gaia2018}, likely leading to a boost in completeness.} The more clearly resolved dip that occurs over the range $10<G<12$ is due to the ``different CCD gates'' \citep{Lindegren2018b}, which we discuss in more detail in Appendix \ref{sec:gating}. This happens well before the switch from $2\times2$ pixel binning to $4\times4$ pixel binning at $G=13$ \citep{Gaia2016} \edits{and so is likely to be unrelated}.  The existence of features like these was always likely given the complexity of the \gaia~\edits{instrumentation}.

An alternative way to visualise the completeness of \gaia is to compute -- for each pixel on the sky -- the faintest magnitude at which \gaia will still see 99\% of the stars at that magnitude. We describe our procedure for computing that magnitude limit in Appendix \ref{sec:completenessAB}. The sky map of the result is shown in the top panel of Fig. \ref{fig:completenessmaps}. We can see that \gaia DR2 is complete down to $G=20$ over almost the entire sky, but in parts of the sky with many scans (the caustics at Ecliptic latitudes of $\pm45^{\circ}$ in particular) the magnitude limit is fainter than $G=21$.

\section{Effect of crowding on our results}
\label{sec:crowding}
The probability that a source is detected by \gaia is lower in highly crowded regions of the sky, because \gaia cannot assign windows to every source on every observation \edits{and some measurements are deleted on-board due a lack of bandwidth to downlink them.} This drop in detection probability decreases the number of sources that have at least five detections, thus causing the completeness of \gaia DR2 to drop in crowded regions. The effective crowding limit for the $G$ photometry and astrometry is $1,050,000\;\mathrm{sources}\;\mathrm{deg}^{-2}$ \citep{Gaia2016}. When \gaia has the option to assign a window to one of two stars, \gaia will always give the brighter star higher priority, and hence crowding acts to decrease the completeness of faint stars relative to bright stars. The windows required for BP/RP photometry and RVS spectroscopy are larger and so have lower source density crowding limits of $750,000\;\mathrm{sources}\;\mathrm{deg}^{-2}$ and $35,000\;\mathrm{sources}\;\mathrm{deg}^{-2}$ respectively, and therefore the effects of crowding will be even more vital to include in these subsets.

In the previous section we ignored the effect of crowding, with two likely consequences. First, the mean detection probability at faint magnitudes will be dragged to lower values by the stars which are missing observations due to crowding, and a portion of the detection probability spread in Model AB can be attributed to crowding changing the effective detection probabilities in regions with high source densities. Second, our selection functions in the previous section are averaged across the entire sky and so should not be applied specifically to regions with only a small range in source density, because the true selection function will deviate significantly. Given the often frustrating correlation between the stellar density of a region and how astrophysically interesting that region is to study, we decided to investigate the effect of crowding on our results.

We attempted to modify our method to fully account for crowding, but were unable to. The major difficulty we encountered is that the effect of crowding should depend on the true density of sources, but we only have access to the observed density of sources in \gaia DR2. A further complication is that crowding should not be a function solely of the density of sources, but rather of the distribution of those sources with magnitude. A proper accounting of crowding is thus beyond the scope of this paper. Nevertheless, in this section, we demonstrate the effect of crowding by fitting our Model T and AB selection functions to ten subsets of \gaia DR2 that are split by the source density in their vicinity.

\begin{figure*}
	\centering
	\includegraphics[width=1.\linewidth,trim=55 80 0 90, clip]{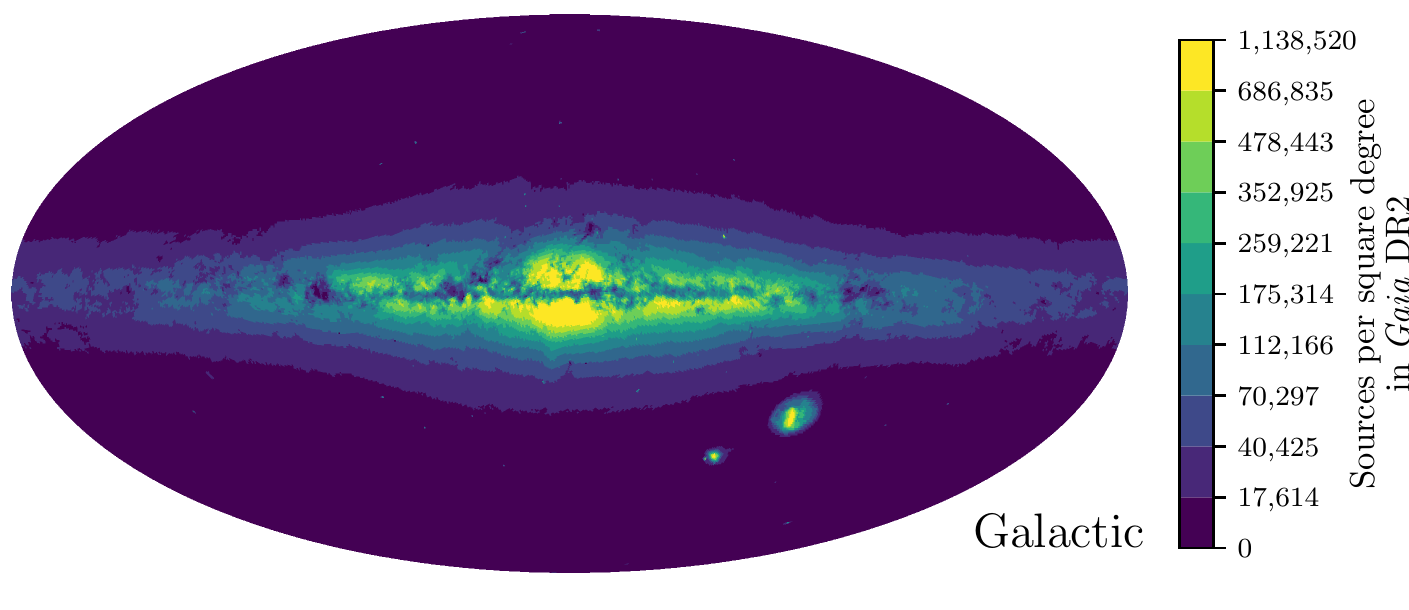}
	\caption{An Equatorial HEALPix map rotated to Galactic coordinates discretely coloured by the number of \gaia DR2 sources per square degree. Note that the edges of the bins in source density were chosen to ensure that an equal number of stars fall into each of the ten bins.}
	\label{fig:skydensity}
\end{figure*}

We split the sky into an \textsc{nside} = 128 Equatorial HEALPix grid, computed the number of \gaia DR2 sources within each pixel, and divided those counts by the $0.21\;\mathrm{deg}^2$ pixel area to obtain the average source density in each pixel. We then grouped the pixels into regions by their source density such that each region contained an equal number of sources, and we show these regions in Fig. \ref{fig:skydensity}. Note that these regions are not contiguous on the sky and that the divisions in source density are irregularly spaced (see the colour bar of Fig. \ref{fig:skydensity} for the divisions). The maximum source density observed in one of these cells of $1,138,520\;\mathrm{sources}\;\mathrm{deg}^{-2}$ is slightly higher than the maximum crowding limit of \gaia, because the sources in a field which are assigned windows changes based on the scan angle.

We modelled the detection probability separately with Models T and AB for each of these pseudo-isodensity regions over the magnitude range \edits{$G < 16$. Below this range} we assumed that the selection function is independent of source density, and thus equal to the selection functions from Sec. \ref{sec:methodology}\edits{, because we should not expect bright stars $G<16$ to be strongly affected by crowding.} We verified the validity of this assumption by fitting Model T to the five least dense and the five most dense regions over the entire magnitude range and comparing the results. The practical motivation for restricting the magnitude range where we fit each region separately is that the dense regions have proportionally fewer bright stars and so the selection function at the bright end is \edits{more poorly} constrained.

\begin{figure*}
	\centering
	\includegraphics[width=1.\linewidth,trim=0 0 0 0, clip]{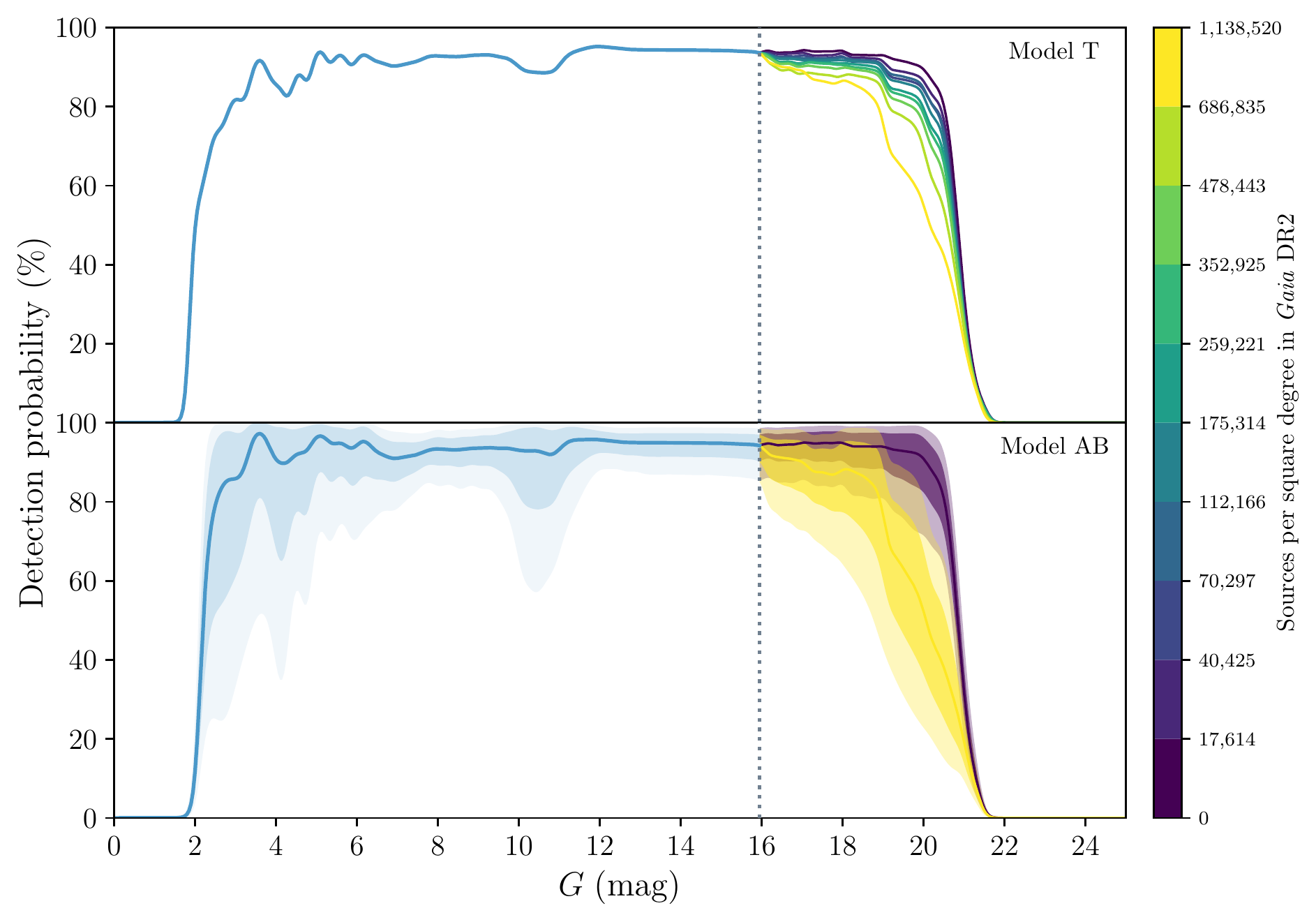}
	\caption{\edits{Our posterior on the detection probability at the median parameters of our} Models T (top) and AB (bottom) when the effect of crowding is included. \edits{In Model T we assume the detection probability is simply a changing function of magnitude, whilst in Model AB we assume it is a changing distribution with magnitude.} In both Models, we assume that crowding only plays a role for \edits{magnitudes $G>16$}. The colour of the lines and regions indicates the pesudo-isodensity region which was used to fit that model. We only show the posteriors for the most and least dense regions in the bottom panel to avoid congesting the plot.}
	\label{fig:crowding}
\end{figure*}

\edits{The median of} our posterior on the detection probability for both models is illustrated in Fig. \ref{fig:crowding}\edits{, where we have discarded the data-points for $G>21.3$ and fitted Gaussian Processes as we did in Sec. \ref{sec:results}.} In Model T, the detection probability consistently decreases with increasing source density, and thus we are able to interpret much of the detection probability spread seen in Fig. \ref{fig:completeness} as due to the varying effect of crowding. We only plotted the Model AB posterior for the least and most dense regions to avoid congesting the figure, but these two distributions bracket the distributions of the other eight regions. The effect of crowding is hugely important, with the $1\sigma$ intervals of the distributions of detection probability in the least and most dense regions being entirely disjoint at $G=20$. The posterior for the least dense region is approximately flat until $G=20$ -- in contrast to the posterior shown in Fig. \ref{fig:modelposterior} -- and is thus representative of the true run of detection probability with magnitude when crowding is not an issue. The posterior for the most dense region exhibits a bump at around $G=18.5$ and we conjecture that this is due to the Red Clump stars in the LMC; this population contains a large number of stars at roughly the same apparent magnitude ($G\approx18.5$) at the same location on the sky, which thus all receive a similar number of observations and so potentially bias the inference at this magnitude. \edits{That the detection probability curves converge on the same behaviour at the faint end is due to the photon-limit becoming more dominant than crowding as a cause of missing detections, although we note that the detection probability of the curve in the densest regions should be further below that in the sparsest regions over the magnitude interval $21.0<G<21.5$. It is possible that there are either spurious sources or sources with miscalibrated magnitudes biasing our determination of the detection probability in this regime that have survived our cuts.}

We show a map of the magnitude limit of 99\% completeness in the bottom panel of Fig. \ref{fig:completenessmaps}, which was computed following the methodology described in Appendix \ref{sec:completenessAB}. The difference in the magnitude limit of 99\% completeness with and without accounting for crowding is striking. In regions with few numbers of observations the magnitude limit has changed by as much as one magnitude. In some extremely dense regions which received few observations \edits{including a large portion of the Galactic bulge}, the magnitude limit is as bright as $G=18$, The competition between numbers of scans and crowding is most visible just West of the Galactic bulge, where the caustics of large numbers of observations push the magnitude limit fainter despite the effect of crowding. The completeness limit in crowded regions will be considerably fainter in later \gaia data releases as the scanning law fills in the gaps.

The explanation for the change in the selection function in regions of few observations between the two panels of Fig. \ref{fig:completenessmaps} is that our inference in Sec. \ref{sec:methodology} was biased by not accounting for crowding. There were many bright stars in crowded regions that had a low detection efficiency which biased Model AB to lower detection probabilities across the entire sky. Accounting for crowding is thus non-negotiable.

\edits{We end this section by reminding the reader that our prescription assumes that the effects of crowding are only conditional on the density of neighbouring sources in \gaia DR2, which is overly simplistic. We also note that we have not accounted for the incompleteness caused by the finite spatial resolution of \gaia, which means that stars in close pairs may not be resolved into separate sources. A further difficulty we have not considered is that bright stars can prevent \gaia from seeing faint neighbouring sources. We will return to these issues in later papers in this series, when we have identified a workable solution.}

\begin{figure*}
	\centering
	\includegraphics[width=1.\linewidth,trim=70 110 35 110, clip]{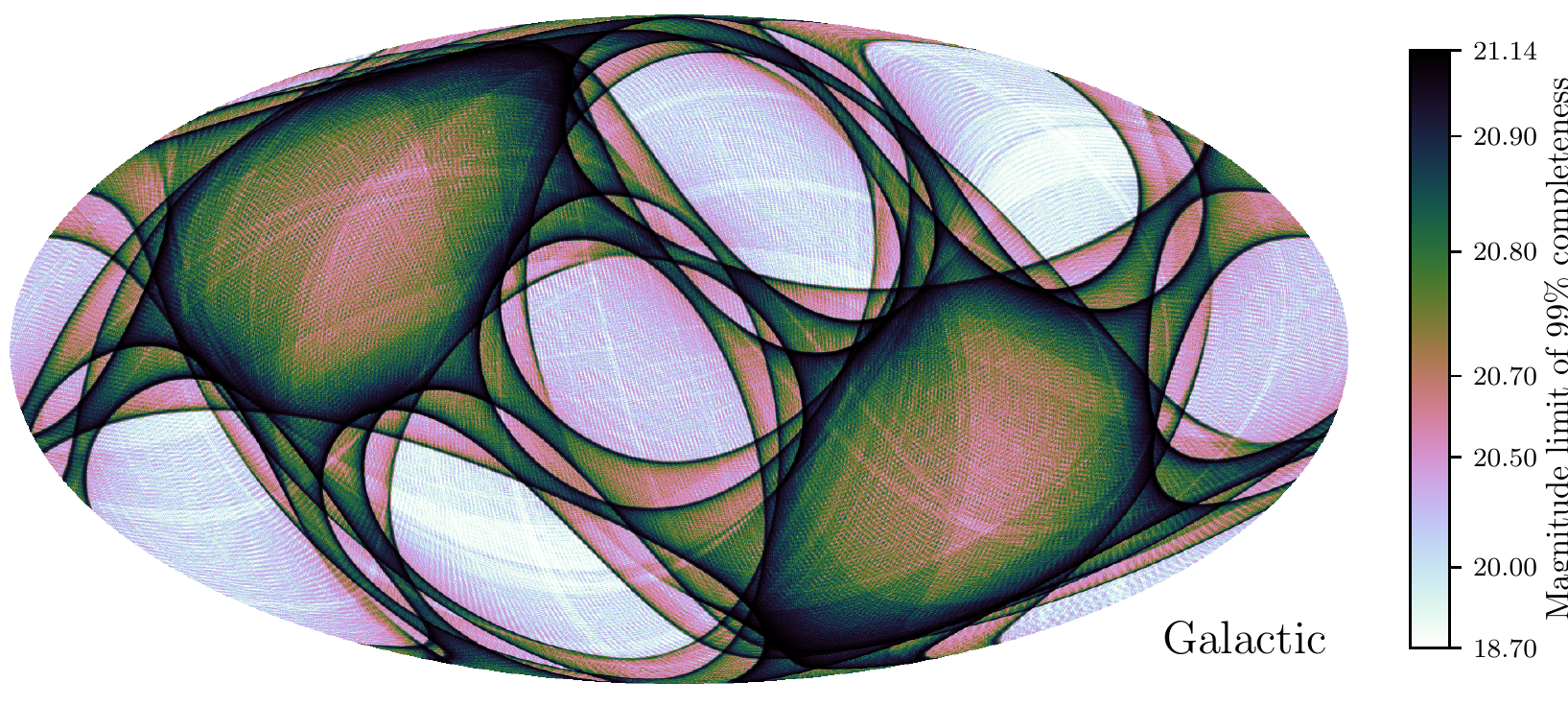}
	(a) \gaia DR2 completeness map based on our 1$^{\mathrm{st}}$-order selection function that only depends on the source magnitude and number of observations.\vspace{0.5cm}
	\includegraphics[width=1.\linewidth,trim=70 110 35 110, clip]{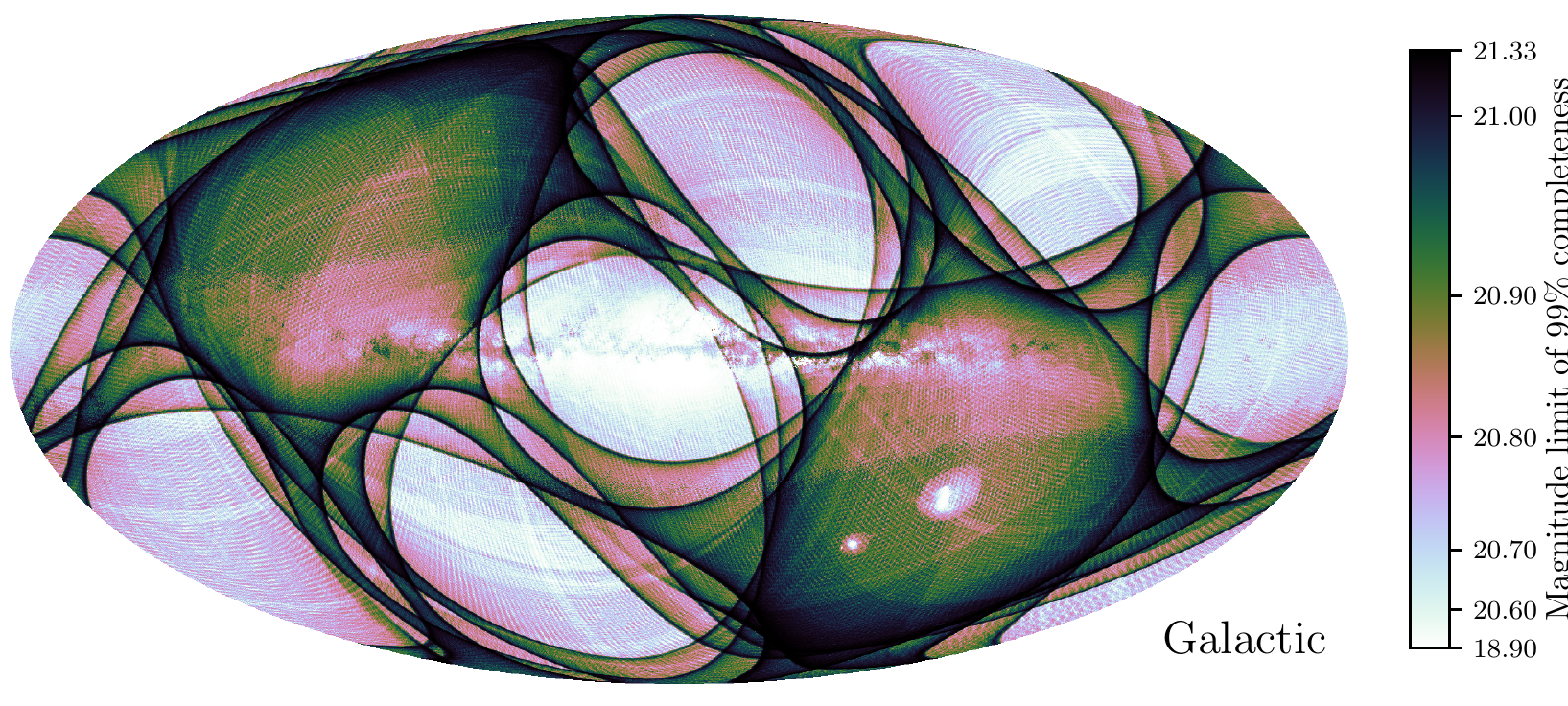}
	(b) \gaia DR2 completeness map based on our 2$^{\mathrm{nd}}$-order selection function that additionally depends on the nearby source density.
	\caption{The magnitude limit down to which \gaia DR2 is 99\% complete varies across the sky due to the scanning law giving different places on the sky different numbers of observations. This limit can be directly calculated from the selection function and scanning law. Here we show HEALPix maps of this magnitude limit assuming either that the selection function only depends on source magnitude and number of observations (top) or that it additionally depends on the \edits{neighbouring} \gaia DR2 source density (bottom). \edits{The colour scale has been histogram-normalized separately in each panel to maximise the dynamic range of the plots in the appropriate region, with the end-points set by the 0.1\% and 99.9\% percentiles of the magnitude limit of the pixels. For the colour map we have used \textsc{cubehelix} \citep{cubehelix}}.}
	\label{fig:completenessmaps}
\end{figure*}

\section{Discussion}
\label{sec:discussion}

\subsection{Previous attempts to quantify the completeness of \gaia DR2}
\label{sec:previous}

\citet{Arenou2018} investigated the completeness of \gaia DR2 as part of the DPAC validation of the catalogue. One approach they used was to calculate the 99$^{\mathrm{th}}$ percentile of $G$ magnitude in pixels on the sky, which they show in their Fig. 3. This definition is different from the magnitude limit we used in Fig. \ref{fig:completenessmaps}, which we defined to be the magnitude at which we predict that \gaia is no longer 99\% complete to stars of that magnitude. \citet{Arenou2018} find that in some parts of the sky their limiting magnitude is fainter than $G=21.5$ and in the Galactic bulge it is as bright as $G=19$, \edits{broadly consistent with our results}.

\edits{A further check carried out by \citet{Arenou2018} was to compare the completeness of \gaia DR2 with that of OGLE \citep{Udalski2008} across a series of fields in the disk, bulge and LMC chosen to have a range of source densities. A limitation of this comparison is the poorer spatial resolution of OGLE compared to \gaia, and so this comparison could only place upper limits on \gaia's completeness. Nevertheless, \citet{Arenou2018} were able to conclude that \gaia is almost complete at $G=18$ across the sky even in high source density regions.}

\edits{In order to improve the test of \gaia's completeness in the densest regions of the sky,} \citet{Arenou2018} \edits{compared} the completeness of \gaia DR2 relative to Hubble Space Telescope observations of 26 globular clusters. In their Fig. 7 they show the completeness as a function of both $G$ magnitude and the density of sources in the field, demonstrating that both of these factors strongly influence the completeness. We note that the completeness of \gaia to stars of the same magnitude in two globular clusters of the same density can be different due to the different number of observations each cluster will have received. This can \edits{likely} explain much of the scatter seen in their Fig. 7.

\citet{Rybizki2018b} estimated the completeness of \gaia DR2 in two ways and made them available through their \textsc{Python} package \textsc{gdr2\_completeness}\footnote{\url{https://github.com/jan-rybizki/gdr2_completeness}}. Their first method divided the sky into a \textsc{nside}=64 HEALPix map and computed the ratio of the number of stars in \gaia DR2 to the number of stars in 2MASS in each of the $G$ magnitude bins $(8,12)$, $(12,15)$ and $(15,18)$ in each of the pixels. This approach will poorly constrain the completeness in pixels with few stars and assumes that the completeness is not varying over large ranges of $G$. It is also limited to the relatively bright magnitude range of the 2MASS catalogue. Their second method assumed that the magnitude at which \gaia DR2 is no longer complete in a pixel on the sky is the mode of the observed magnitude distribution. This approach only returns the true completeness magnitude limit if the true magnitude number density is monotonically increasing, \gaia is 100\% complete up to the limit and 0\% complete beyond it. In general, this is a biased estimator that performs worse in pixels on the sky with few stars. However, their method demonstrates that \gaia drops in completeness in low extinction regions close to the Galactic center where there are high densities of bright sources.

\subsection{Applications in the \gaia-verse}
In the previous sections we demonstrated that \gaia is mostly complete to sources with magnitudes in the range $3<G<20$ (aside from in the densest regions of the sky), however there are further scanning-law-driven selections that can cause sources in the \gaia DR2 source catalogue to not have published parallaxes, proper motions, variability indicators or radial velocities.

\subsubsection{Stars with proper motions and parallaxes}
As mentioned in Sec. \ref{sec:methodology}, more stringent cuts were applied by \citet{Lindegren2018} when determining whether a source has a reported parallax and proper motion. In addition to cuts on a proxy for the size of the astrometric uncertainty and on $G<21$, \citet{Lindegren2018} cut to require that there were at least 6 clusters (\textsc{visibility\_periods\_used}) amongst the $k$ detections, as this ensured a reasonable spread of the measurements across the time period of \gaia observations. These additional cuts are challenging to interpret as a selection function and so we postpone their dissection to \edits{a later} paper of this series.

\subsubsection{Variable stars}
One of the major data products associated with \gaia DR2 were the catalogues of classified variable stars \citep{Mowlavi2018, Clementini2019}. The pipeline for these classifications discarded some fraction of the observations of each star as failing to meet their quality criteria, and thus defined $k_{\mathrm{good}}\leq k \leq n$ as the number of good observations of each star. Cuts were then made on $k_{\mathrm{good}}$ at different points in the variability pipeline (see Fig. 2 of \citealp{Holl2018}). The weakest of these cuts was that a source was only classified as variable or non-variable (\textsc{phot\_variable\_flag} in the \gaia DR2 main catalogue) if $k_{\mathrm{good}}\geq2$, while sources were only classified as RR Lyrae, Cepheids or Long Period Variables if $k_{\mathrm{good}}\geq12$ and as Short Period or Rotationally Modulated Variables if $k_{\mathrm{good}}\geq20$. For comparison, 826 million stars satisfy $k_{\mathrm{good}}\geq20$ compared to 833 million stars that satisfy $k\geq20$ and 1,124 million stars that satisfy $n\geq20$. These numbers suggest that the observation quality cuts applied by the DPAC are not removing many detection and thus any selection of the form $k_{\mathrm{good}}\geq K$ can be approximated by the selection $k\geq K$. This approximation is necessary because the quantity $k_{\mathrm{good}}$ was only published for certain subsets of the variability catalogues. The selection function of a specific variable star catalogue in \gaia DR2 (those stars that are classified as being a variable of a specific type) is likely to have at least four components, which we will illustrate for the Rotationally Modulated Variables (otherwise known as BY Dra-type stars):
\begin{enumerate}
    \item Was the star observed at least $n\geq20$ times?
    \item Was the star detected at least $k\geq20$ times?
    \item Were at least $k_{\mathrm{good}}\geq20$ of those detections of good quality?
    \item Does supervised classification based on the location of the star in the H-R diagram and the time series of photometric measurements from those good detections support the conclusion that this is a BY Dra-type star?
\end{enumerate}
The first three of these selections try to select only those stars where DPAC were confident in being able to reliably classify the variability, while the final selection attempts to select those stars that are actually BY Dra-type variables. 

Curiously, the sky map of the number of \gaia-classified BY Dra-type variables (see Fig. 6 of \citealp{Holl2018}) suggests that the effective number of good detections required for the classification was much higher than 20, because the classified stars are only found along the caustics of many repeated scans and in regions covered by the Ecliptic Pole Scanning Law (EPSL). Strikingly -- apart from the EPSL -- the North and South Ecliptic polar cap regions, which typically have a quite high number of observations of $30<n<40$, are entirely devoid of BY Dra stars. \citet{Lanzafame2018} attempted to quantify the completeness of the \gaia BY Dra catalogue by extrapolating the $14\%$ completeness of their catalogue in the region around the Pleiades (in comparison to an existing catalogue) to the 38\% of the sky in which their catalogue contains BY Dra stars, and thus arriving at an upper limit of 5\% completeness. Comparing Fig. A.1. of \citet{Lanzafame2018} to our Fig. \ref{fig:map}, we can see that BY Dra stars appear to only be detected on the two caustics of between forty and fifty observations that cross through the Pleiades. It is clear that the reported 14\% completeness is being driven by the number of observations that part of the sky received, and thus that this number should not be extrapolated to the rest of the sky. Indeed, a large fraction of the 38\% of the sky containing \gaia BY Dra stars has more than fifty observations, and thus should be more than 14\% complete. In Fig. 6 of \citet{Holl2018}, the number of detected BY Dra variables increases towards the South Ecliptic Pole even as the true number density must be decreasing (due to increasing Galactic latitude), which affirms our conclusion that the completeness of the \gaia BY Dra variable catalogue is being strongly determined by the number of observations.

\edits{A further complication is that the ability of DPAC processing to classify a source as a variable will depend on the timings of the detections, not just their quantity. For instance, sources near the Ecliptic Poles will have received a large number of observations during the month-long EPSL, but those many observations will not be as constraining of the period of a Long Period variable star as a few observations evenly spaced across the 22 months of DR2.}

\gaia is revolutionary for the study of variable stars, because it is the first all-sky mission with large numbers of repeat visits with high photometric precision down to $G=20$ covering multi-year baselines. An accurate estimate of the completeness of the \gaia variable star catalogues will need to account for the unique spinning-and-precessing way that \gaia looks at the sky. While this is outside the scope of this paper, we plan to return to this topic in later papers in this series.

\subsubsection{Stars with radial velocities}
\gaia DR2 included the largest spectroscocopic radial velocity catalogue in history, with the on-board Radial Velocity Spectrometer (RVS) reporting radial velocities of 7,224,631 stars \citep{Cropper2018,Sartoretti2018,Katz2019}. The quantity \textsc{radial\_velocity} in \gaia DR2 is the median of the multiple $k_{\mathrm{RVS}}$ radial velocity measurements \gaia made of each star. DPAC required that a star had at least $k_{\mathrm{RVS}}\geq2$ measurements in order for the star to be published in \gaia DR2. This selection has the effect of making the \gaia DR2 RVS subset more complete in regions of the sky with more observations, as can be seen in Fig. 6 of \citet{Katz2019}. \gaia attempts to measure the radial velocity of a star if all of the following apply: 1) the star was detected by the SkyMapper CCDs, 2) the star is crossing the focal plane in one of the four of seven rows which have an RVS CCD, and 3) \gaia was able to assign a window to the star that does not partially overlap with a window already assigned to another star. The first requirement is not onerous because the \gaia DR2 RVS catalogue was limited to magnitudes $G_{\mathrm{RVS}}<12$ and \gaia detects \edits{most} stars this bright on almost every observation. The second requirement \edits{can be thought of as discarding} three out of every seven observations, and thus if we consider $k_{\mathrm{RVS}}\sim\operatorname{Bin}(n,\theta_{\mathrm{RVS}})$ then $\theta_{\mathrm{RVS}}\leq 4/7$. The third requirement greatly complicates the $k_{\mathrm{RVS}}\geq2$ selection, because the odds that there is already a star with an assigned window that would overlap are dramatically higher in denser regions of the sky. This explains the drop in completeness of \gaia RVS near the Galactic plane, and the particularly low 25\% completeness near the Galactic centre where the high density of sources combines with a region of small numbers of observations. Whether the windows of two sources will overlap depends on the angle that \gaia is scanning across that field, and thus changes between each scan. \edits{A further complication is that whether a radial velocity measurement can be extracted from the RVS spectrum will depend on the depth of the Calcium triplet which will vary with the surface gravity and effective temperature of the source.} We therefore conjecture that the probability that an observation of a star will result in a successful radial velocity measurement will be a \edits{complicated} function of the magnitude $G_{\mathrm{RVS}}$, the colour $G_{\mathrm{BP}}-G_{\mathrm{RP}}$ \edits{(acting as a proxy for the stellar type)} and the density of nearby bright sources.

The \gaia DR2 catalogue of radial velocities has been transformative for Galactic dynamics, but sophisticated applications have been hindered by the lack of a well-motivated selection function that accounts for the scanning law selection. The development of such a selection will be the subject of two later papers in this series.

\subsection{Poisson binomial distribution}
\label{sec:poissonbinomial}
One of the key assumptions of our methodology is that the detection probability $\theta$ of each source is the same every time that source is observed (see Sec. \ref{sec:framework}). There are several possible ways that this assumption could be violated in reality, for instance:
\begin{enumerate}
	\item If the source is variable, then the detection probability will vary depending on the brightness of the star at the time of each observation.
	\item If the source is faint and in a crowded region, then whether the source can be assigned a non-overlapping window can depend on the angle at which \gaia is scanning across the field, which changes between scans.
	\item \gaia suffers from stray light being scattered by fibres at the edge of the sunshield \citep{Gaia2016}. The level of straylight contamination varies with the location of \gaia on its orbit, thus causing a time-dependence in the detection probability of faint sources.
\end{enumerate}

A possible improvement of our method would be to assume that the detection probability of a source is different on each observation, and thus that the number of detections is a Poisson-Binomial random variable. Suppose there were $n$ observations with detection probabilities $\{p_1,\dots p_n\}$. The probability of $k$ detections is then given by
\begin{equation}
\operatorname{P}(k|\{p_i\}) = \sum_{A\in F_k^n} \prod_{i\in A} p_i \prod_{j\in A^{\mathrm{c}}} (1-p_j),
\end{equation}
where $F_k^n$ is the set of all subsets of $k$ integers that can be picked from $\{1,\dots,n\}$. The Poisson-Binomial \edits{probability mass function} is computationally-challenging to evaluate for even moderately large $n\gtrsim20$, because the size of the set $F_k^n$ -- and thus the number of terms to be summed -- is $|F_k^n|=\binom{n}{k}$.  However, if we assume that each $p_i\sim\operatorname{Beta}(\alpha,\beta)$, where we assume that the $\alpha$ and $\beta$ are the same across all observations of a single source, then we can marginalise away the $p_i$ by noting that
\begin{align}
\operatorname{P}(k&|\alpha,\beta)=\idotsint \operatorname{P}(k|\{p_i\})\prod_{j=1}^n\operatorname{P}(p_j|\alpha,\beta) \mathrm{d}p_j \nonumber\\
&= \sum_{A\in F_k^n} \prod_{i\in A} \int p_i \operatorname{B}_{\alpha,\beta}(p_i) \mathrm{d}p_i \prod_{j\in A^{\mathrm{c}}} \int (1-p_j) \operatorname{B}_{\alpha,\beta}(p_j) \mathrm{d}p_j \nonumber \\
&= \sum_{A\in F_k^n} \prod_{i\in A} \frac{\operatorname{B}(\alpha+1,\beta)}{\operatorname{B}(\alpha,\beta)} \prod_{j\in A^{\mathrm{c}}} \frac{\operatorname{B}(\alpha,\beta+1)}{\operatorname{B}(\alpha,\beta)} \nonumber\\
&= \sum_{A\in F_k^n} \prod_{i\in A} \frac{\alpha}{\alpha+\beta} \prod_{j\in A^{\mathrm{c}}} \frac{\beta}{\alpha+\beta} \nonumber\\
&= \sum_{A\in F_k^n} \left(\frac{\alpha}{\alpha+\beta} \right)^{|A|}\left(\frac{\beta}{\alpha+\beta} \right)^{|A^{\mathrm{c}}|} \nonumber\\
&= \binom{n}{k} \left(\frac{\alpha}{\alpha+\beta} \right)^{k} \left(1-\frac{\alpha}{\alpha+\beta} \right)^{n-k}.
\label{eq:pbin}
\end{align}
This is the form of the probability mass function of a Binomial random variable, and thus we can equivalently write that $k\sim\operatorname{Binomial}(n,\frac{\alpha}{\alpha+\beta})$. We note that the combination $\frac{\alpha}{\alpha+\beta}$ is the mean of the $\operatorname{Beta}(\alpha,\beta)$ distribution, which implies that -- when the $p_i\sim\operatorname{Beta}(\alpha,\beta)$ -- the number of detections $k$ is only sensitive to the ratio of $\alpha$ and $\beta$ and not their scale.

By marginalising the $\{p_i\}$ we have drastically simplified the evaluation of the Poisson-Binomial likelihood with Beta priors, and therefore enabled it for use with large $n$. However, this simplification can only be made if the data $k$ has not been truncated. If the data has been truncated $k\geq5$, then the likelihood must be re-normalised by the survival function,
\begin{equation}
\operatorname{S}(k\geq5|\{p_i\}) = 1- \sum_{l=0}^4\sum_{A\in F_l^n} \prod_{i\in A} p_i \prod_{j\in A^{\mathrm{c}}} (1-p_j),
\end{equation}
and thus this intimidating function would appear as a divisor inside the integral in the derivation above. It is possible that there exists a closed-form solution to this integral, but, if it exists, then it is likely to be in terms of obscure special functions. We plan to revisit the possibility of utilising a truncated Beta-Poisson-Binomial model \edits{in a later paper in this series}.

\subsection{Using our selection functions}
To aid the reader in using our selection functions, we have created a new \textsc{Python} module \textsc{selectionfunctions} (\url{https://github.com/gaiaverse/selectionfunctions}) based on the \textsc{dustmaps} package by \citet{Green2018}. This module allows the user to easily query our selection functions in any coordinate system. We have shown in Sec. \ref{sec:crowding} that when crowding is not accounted for the selection function is biased and we therefore do not provide the selection functions computed in Sec. \ref{sec:methodology} in \textsc{selectionfunctions}. However, we provide the ability to query both Model T and Model AB from Sec. \ref{sec:crowding}, and to ignore the effect of crowding by only returning the selection function computed for the least dense bin (the purple lines in Fig. \ref{fig:crowding}). We illustrate the simplicity of using our \textsc{selectionfunctions} package in the code snippet below, where we query our Model AB from Sec. \ref{sec:crowding} to calculate the selection function for the fastest main-sequence star in the Galaxy \citep[S5-HVS1,][]{Koposov2020}. \edits{In the future, we plan to include the selection functions of ground-based spectroscopic surveys \citep{Everall2020} and the ability to query the selection function of the intersection of multiple surveys, thus enabling astronomers to ask questions such as `what are the odds that a star in \gaia DR2 has an APOGEE radial velocity?'.}

\begin{lstlisting}[language=Python]
import selectionfunctions.cog_ii as cog_ii
from selectionfunctions.source import Source

dr2_sf = cog_ii.dr2_sf(version='modelAB')
s5_hvs1 = Source('22h54m51.68s',
           '-51d11m44.19s',
           photometry={'gaia_g':16.02},
           frame='icrs')
print(gaia_dr2(s5_hvs1))

\end{lstlisting}

\section{Conclusions}
\label{sec:conclusion}

The \gaia mission has broken astrometric, photometric and spectroscopic records with its second data release, but to \edits{fully exploit this remarkable dataset we need to know which stars are missing and where they lie on the sky. In this work we argued that the completeness of \gaia DR2 is driven by the spinning-and-precessing way that \gaia looks at the sky.} We computed the number of times that \gaia looked at each of the stars in \gaia DR2 and at each point on the sky. We then developed a statistical framework with which we answered the question: what are the odds that \gaia detects a star each time it observes it? The answer to this question is important, because to leading order \gaia DR2 contains all sources that were detected at least five times. We could answer this question because \gaia reports the number of times that each source was detected. We modelled the run of the detection probability with $G$-band magnitude, finding that \gaia DR2 is broadly complete over $7<G<20$ but that the completeness falls to 0\% over the range \edits{$20<G<21.3$}. We created an interactive visualisation to illustrate this result. We further calculated the magnitude up to which \gaia is complete at each location on the sky, finding that this magnitude limit varies from 18.0 to 21.3. We then extended our selection function to account for crowding, and found that crowding is a vitally important component of the \gaia selection function, particularly in the Galactic bulge \edits{and the Large and Small Magellanic Clouds.} We concluded by conjecturing on the necessity of accounting for the \gaia scanning law when modelling the selection functions of the \gaia DR2 parallax and proper motion, variable star and radial velocity sub-catalogues, topics that we will return to in later papers of this series. \edits{Finally, we presented a \textsc{Python} package that allows the reader to easily incorporate our selection functions in their work.} This work lays down the conceptual framework needed to grasp the selection functions of the \gaia catalogues, and thus is a fundamental stepping stone on the path to mapping our Galaxy with \gaia.

\section*{Acknowledgements}
The authors are grateful to the \gaia DPAC for making the scanning law publicly available and to Berry Holl for his help in interpreting it. DB thanks Magdalen College for his fellowship and the Rudolf Peierls Centre for Theoretical Physics for providing office space and travel funds. AE thanks the Science and Technology Facilities Council of
the United Kingdom for financial support. This work has made use of data from the European Space Agency (ESA) mission \gaia (\url{https://www.cosmos.esa.int/gaia}), processed by the \gaia
Data Processing and Analysis Consortium (DPAC,
\url{https://www.cosmos.esa.int/web/gaia/dpac/consortium}). Funding for the DPAC
has been provided by national institutions, in particular the institutions
participating in the \gaia Multilateral Agreement.




\bibliographystyle{mnras}
\bibliography{references} 




\appendix


\section{Intuition for Binomial and Beta distributions}
\label{sec:betabinomial}

\begin{figure}
	\centering
	\includegraphics[width=1.\linewidth,trim=5 20 0 5, clip]{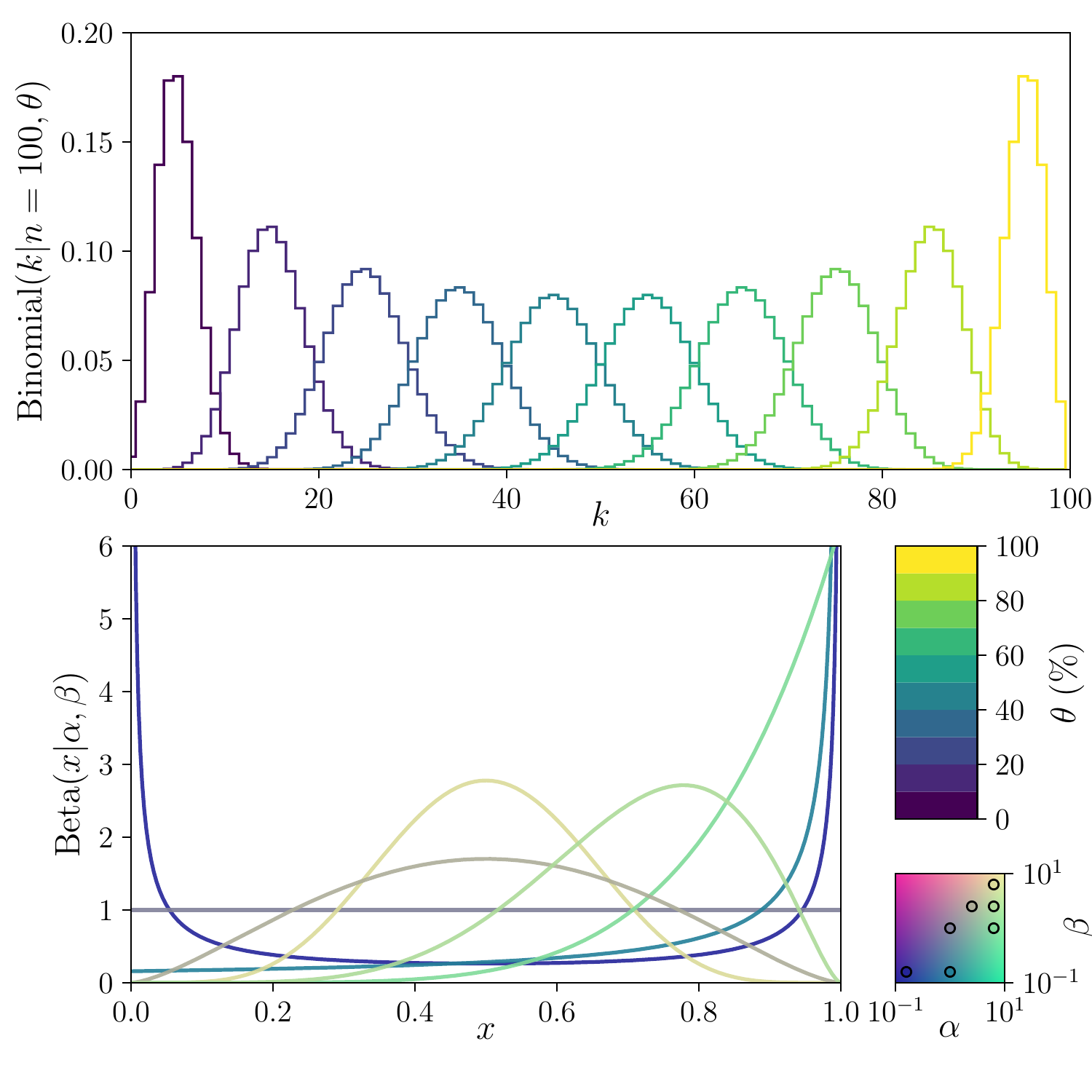}
	\caption{Illustrations of the Binomial and Beta distributions. \textbf{Top:} The probability mass function of a $\mathrm{Binomial}(n,\theta)$ random variable where $n=100$ and $\theta$ is varied from 5\% to 95\% in steps of 10\%. \textbf{Bottom:} The probability density function of a $\mathrm{Beta}(\alpha,\beta)$ random variable for seven different choices of $\alpha$ and $\beta$. The colour of the lines is logarithmically mapped to the bicolour map in the bottom right, where black circles correspond to the values of each line.}
	\label{fig:betabinomial}
\end{figure}

We have included this appendix to introduce the reader to the shapes of the Binomial and Beta distributions, which are not commonly used in astronomy\footnote{The term \textit{binomial} appeared in the full texts of 370 astronomy papers in 2018 according to the SAO/NASA Astrophysics Data System, as compared to 10,874 astronomy papers which mentioned the term \textit{gaussian}.}. We illustrate in Fig. \ref{fig:betabinomial} the variety of probability mass functions of a Binomial random variable (top) and the probability density functions of a Beta random variable (bottom) that are possible. The Binomial distribution has only one shape parameter $\theta$ and thus the behaviour is quite simple. The Beta distribution, however, has two shape parameters and so can exhibit a wide range of shapes. We only show the PDF for values of $(\alpha,\beta)$ where $\alpha\geq\beta$, because swapping the values of $\alpha$ and $\beta$ simply reflects the PDF about $x=0.5$. We highlight that if $\alpha,\beta\rightarrow\infty$ with $\alpha/\beta$ fixed then the Beta distribution tends towards a \edits{bell-curve} centred on $\alpha/(\alpha+\beta)$.

\section{Deconvolution to account for duplicate detections}
\label{sec:deconvolution}

\begin{figure}
	\centering
	\includegraphics[width=1.\linewidth,trim=0 0 0 0, clip]{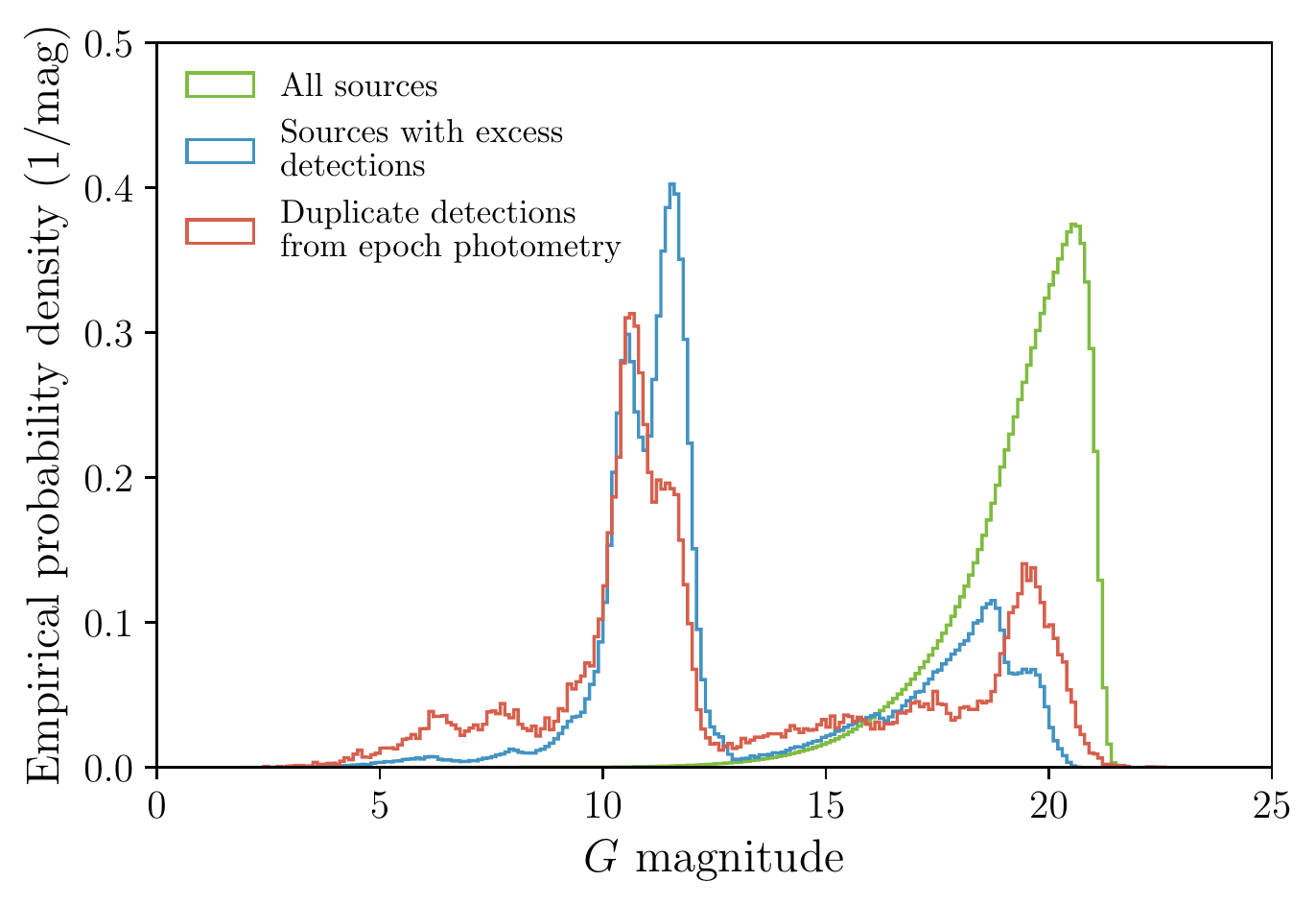}
	a) Magnitude distributions of sources with excess detections.\vspace{0.5cm}
	
	\includegraphics[width=1.\linewidth,trim=0 0 0 0, clip]{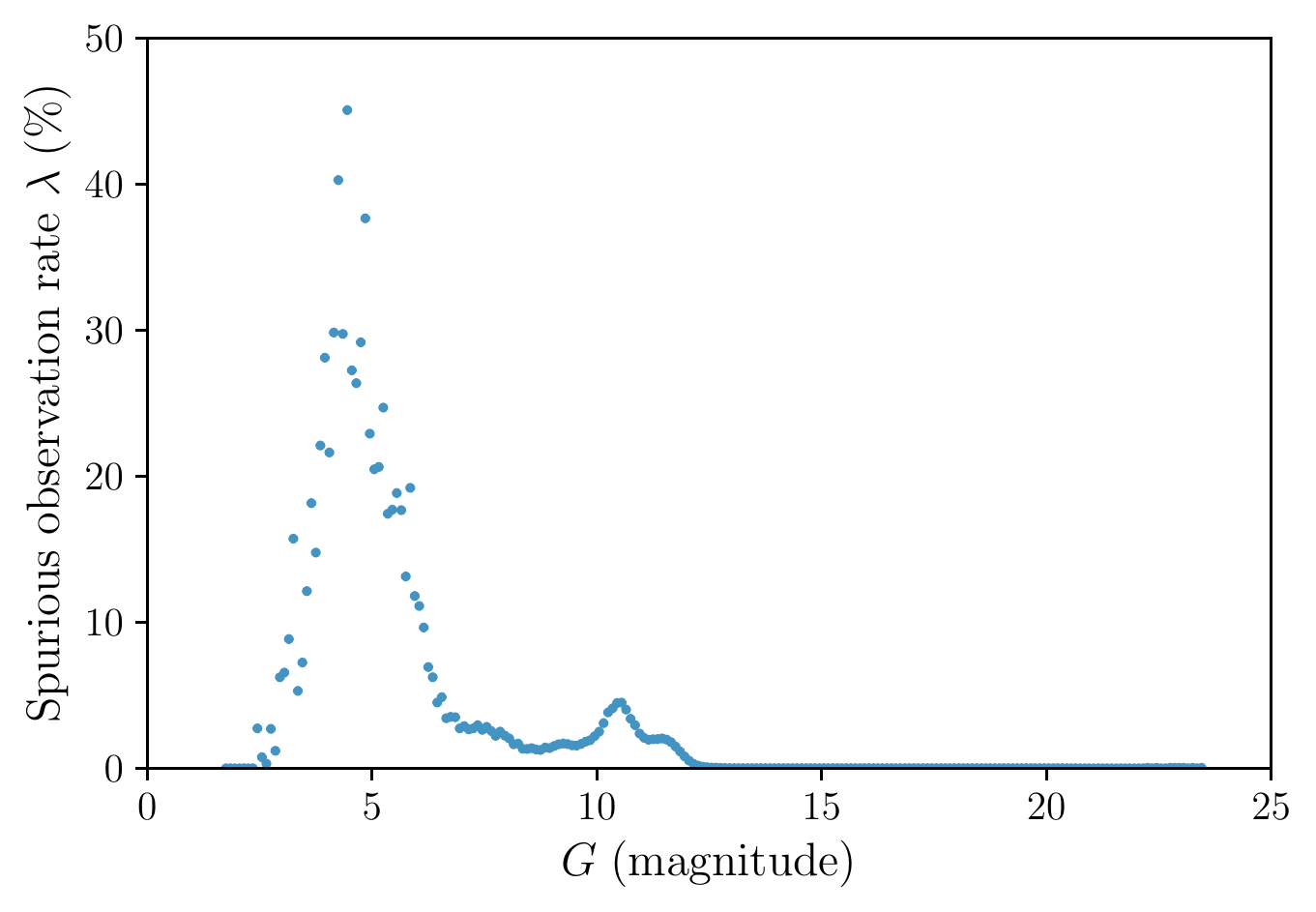}
	b) Estimate of the spurious observation rate.\vspace{0.5cm}
	
	\caption{\edits{The paradox of sources with more reported astrometric detections than predicted observations can be explained by spurious duplicate detections of bright sources and small $(\pm1)$ errors in our prediction of the number of observations. \textbf{Top:} The magnitude distribution of sources with excess detections (blue) is not consistent with the magnitude distribution of all sources in \gaia DR2 (green), instead bearing a strong resemblance to the magnitude distribution of spurious duplicate \gaia detections of variable stars that we identified in Paper I \citep{Boubert2020}. \textbf{Bottom:} The methodology described in the main text requires the number of detections $k$ to be less than or equal to the number of observations $n$. We deconvolved the distribution of stars in $(n,k)$ as a function of $G$ to ensure this was true. A by-product of this was an estimate of the spurious observation rate $\lambda$, e.g. the fraction of true observations of the source that could have resulted in a spurious detection.}}
	\label{fig:excess}
\end{figure}

\begin{figure}
	\centering
	
	\includegraphics[width=1.\linewidth,trim=0 0 0 0, clip]{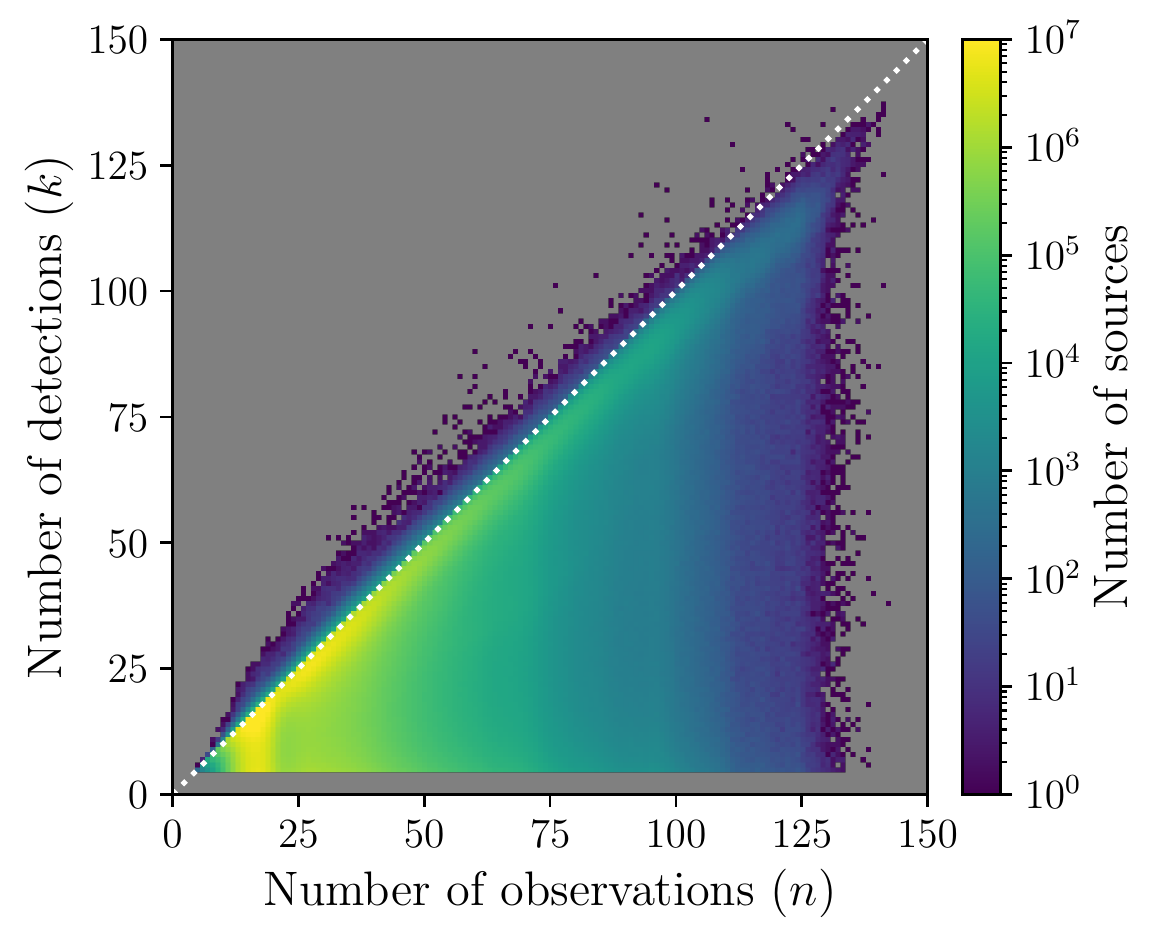}
	a) Distribution of detections and observations before deconvolution.\vspace{0.5cm}
	
	\includegraphics[width=1.\linewidth,trim=0 0 0 0, clip]{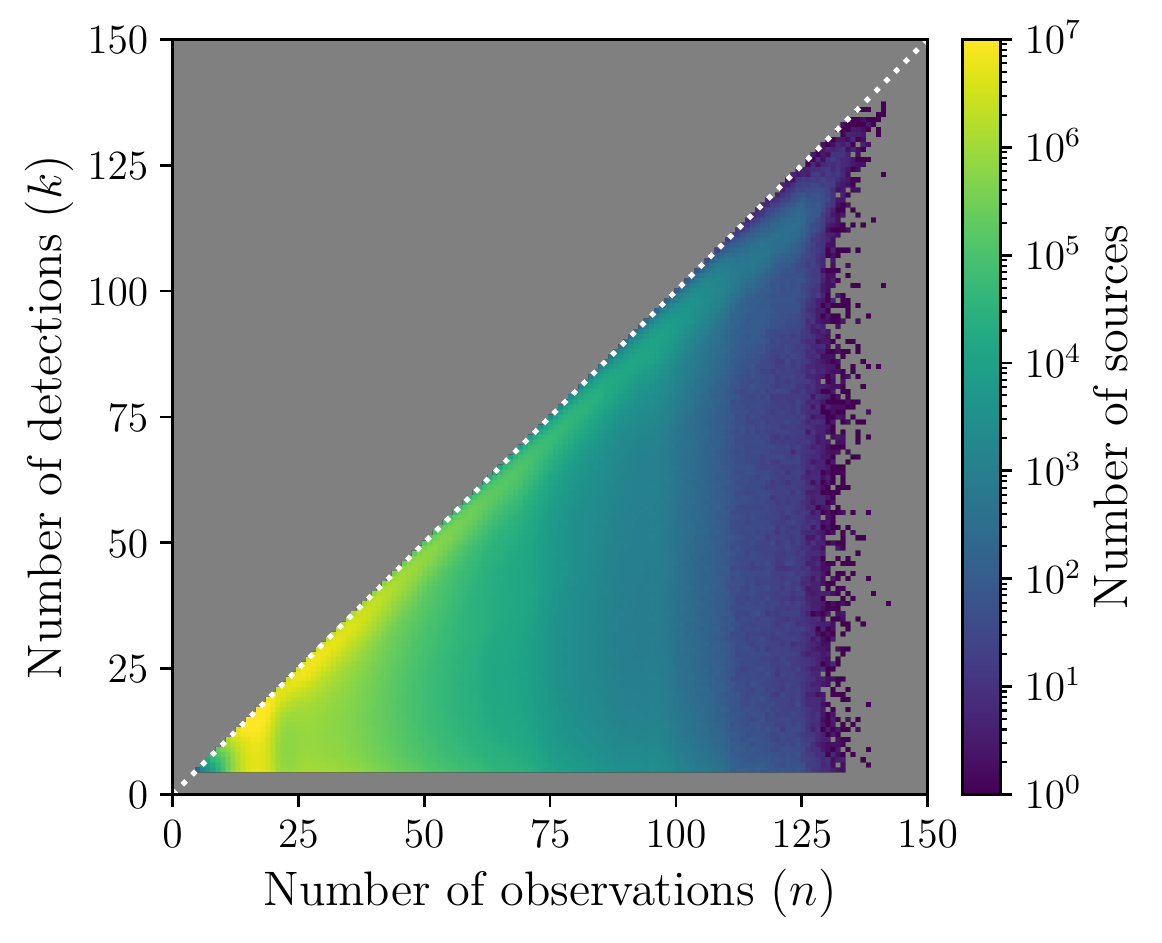}
	b) Distribution of detections and observations after deconvolution.\vspace{0.5cm}
	
	\caption{\edits{The distribution of all sources in \gaia DR2 (excluding those cut due to suspected miscalibrated magnitudes) by their number of reported astrometric detections and predicted observations before (top) and after (bottom) we applied our deconvolution methodology.}}
	\label{fig:duplicates}
\end{figure}

\edits{After predicting the number of observations of every source in \gaia DR2, we binned the sources by their predicted number of observations $\tilde{n}$, reported number of detections $k$ and $G$ magnitude. Paradoxically, there are 619,272 sources with more reported detections than predicted observations ($k>\tilde{n}$). We are aware of two possible reasons that this could occur. Firstly, the predicted number of observations $\tilde{n}$ may differ from the true number of observations $n$ due to small inaccuracies in our methodology. For instance, we are using the scanning law we derived in Paper I from the light curves of the DR2 variable stars for which the locations of the fields of view are only accurate to $66\;\mu\mathrm{as}$. This limitation -- while small -- could cause some stars to lose out on observations whilst others could gain observations; crucially, this effect is purely geometric and will be independent of the magnitude of the source. Secondly, some sources can have spurious duplicate detections (as discussed in detail in Paper I), where more than one detection is attributed to a source during a single field-of-view transit and which is known to occur more frequently for bright stars. In the top panel of Fig. \ref{fig:excess} we show the empirical magnitude density distribution of the sources with excess detections ($k>\tilde{n}$), in addition to the empirical magnitude density distribution of the spurious duplicate detections we identified in Paper I from the epoch photometry of the \gaia DR2 variable stars. The similarity between these two distributions (and the dissimilarity with the magnitude distribution of all \gaia DR2 sources shown for reference) suggests that the phenomenon of sources with excess detections is at least partially due to spurious duplicate detections.}

\edits{The statistical methodology discussed in the main text cannot be used if there are more detections than observations, because you cannot flip a coin five times and see six heads. We could simply discard any sources with $k>\tilde{n}$, however this could lead to a bias in our final results. Our solution was to instead introduce $N$, the \textit{effective} number of observations which we define to be the sum of the true number of observations $n$ and the number of occasions on which a spurious detection could have occurred $m$. We are therefore interpreting the spurious duplicate detections as additional bonus observations of the source.} 

\edits{Suppose for a source with magnitude $G$ that the true number of observations was $n$ and the reported number of detections was $k$. We assume that the predicted number of observations $\tilde{n}$ is a random draw from a Normal distribution $\tilde{n}\sim\operatorname{Normal}(n,\sigma)$, where $\sigma$ is the spread of our observation prediction error. We further assume that the true number of duplicate observations $m$ (occasions on which a spurious detection may occur) is distributed like a Poisson random variable with a rate that is proportional to the number of true observations $m\sim\operatorname{Poisson}(\lambda n)$, where $\lambda=\lambda(G)$ is a free parameter that can be interpreted as the fraction of true observations of a source of magnitude $G$ that are accompanied by a duplicate observation. The effective number of observations is then simply $N\equiv n + m$. As we noted in Paper I, these additional spurious detections are counted towards the five detection criterion used to select stars for \gaia DR2. The observed distribution of sources in $(G,\tilde{n},k)$ is then a convolution of $(G,N,k)$ with the normal and Poisson distributions defined above.}

\edits{We have the number of stars $I(G,\tilde{n},k)$ in bins of $G$, $\tilde{n}$ and $k$. We wish to obtain an estimate of the number of stars $J(G,N,k)$ in bins of $G$, $N$ and $k$. Given $\sigma$ and $\lambda(G)$, we can define a transition matrix $M_{G,k}(\tilde{n},N|\sigma,\lambda)$ that gives the fraction of the stars in bin $(G,N,k)$ prior to the convolution that end up in bin $(G,\tilde{n},k)$ after the convolution. We employ the Richardson-Lucy algorithm \citep{Richardson1972,Lucy1974} which -- for a given convolved image and the transition matrix -- iteratively converges on a maximum likelihood estimate of the deconvolved image. We do not know the correct values of $\sigma$ and $\lambda(G)$ in advance and so obtain them through a second layer of maximum likelihood optimisation, where we use a Poisson likelihood with a rate in each bin given by the convolution of the maximum likelihood deconvolved image, i.e. we compute how likely  $I(G,\tilde{n},k)$ is to be the result of convolving  $M_{G,k}(\tilde{n},N|\sigma,\lambda)$ with $J(G,N,k)$. We enforced the requirement that $J(G,N,k)=0$ if $N<k$ by forcing the transition matrix to send all stars in those bins in the deconvolved image to $I(G,0,k)$, and -- because $I(G,0,k)$ is always zero due to the \gaia selection cut -- the Richardson-Lucy algorithm will always favour sending  $J(G,N,k)$ to zero if $N<k$. We have no intuition for a particular functional form for $\lambda(G)$ and so we modelled it as piece-wise constant, such that it has a different value in each $G$ bin. We adopted log-uniform priors on $\sigma$ and $\lambda(G)$. To make the problem entirely disjoint between each slice of $G$ (and thus simplify the computation), we let $\sigma$ vary between $G$ slices and so could optimise $(\sigma,\lambda)$ independently in each slice of $G$. We found that $\lambda$ assumed a negligible value for $14<G<20$ whilst $\sigma$ was consistent with being constant in that regime $\sigma = 0.0625\pm 0.0022$. For $G<14$ we found that $\lambda$ took values as large as 0.4 (implying that 40\% of observations could have been accompanied by a spurious duplicate detection), whilst $\sigma$ varied and dipped as low as $\sigma=0.02$, suggesting some degeneracy between $\sigma$ and $\lambda$. We opted to fix $\sigma= 0.0625$ and then re-ran the -- now truly disjoint -- optimisation in each $G$ slice. In the bottom panel of Fig. \ref{fig:excess} we show the resulting maximum likelihood estimate of the rate of observations that could have resulted in spurious detections. It shows the same peak in the range $10<G<12$ that is visible in the top panel, but the greatest rate occurs around $G = 5$. We note that most of the sources with more than five excess detections do lie in a peak in this bright magnitude range, as shown in Fig. 9b of Paper I where we performed a similar excess detection analysis when investigating the spurious duplicate detection phenomenon.} 

\edits{The result of these deconvolutions are shown in Fig. \ref{fig:duplicates} where in the top panel we show the original distribution $I(G,\tilde{n},k)$ and in the bottom panel we show our maximum likelihood estimate of $J(G,N,k)$, with both distributions being summed over $G$. The deconvolutions have not substantially altered the structures visible in this plot, which we should expect given that almost all of the stars are dimmer than $G=14$ and so were only deconvolved with respect to a Gaussian of width 0.0625. We note that the prediction error likely has heavier-than-Gaussian tails and that in Paper I we estimated our observation prediction error using a Skellam distribution for this reason, but the probability mass function of the Skellam is in terms of a modified Bessel function of the first kind and so would have been computationally intensive to use in this application. We also applied our deconvolution methodology to the grids used in Sec. \ref{sec:crowding}, but we fixed both $\sigma$ and $\lambda(G)$ to the maximum likelihood values obtained above. In the remainder of the text we will refer to both $\tilde{n}$ and $N$ as $n$ to aid clarity.}



\section{Determining completeness for Model AB}
\label{sec:completenessAB}

\begin{figure}
	\centering
	\includegraphics[width=1.\linewidth,trim=5 25 65 25, clip]{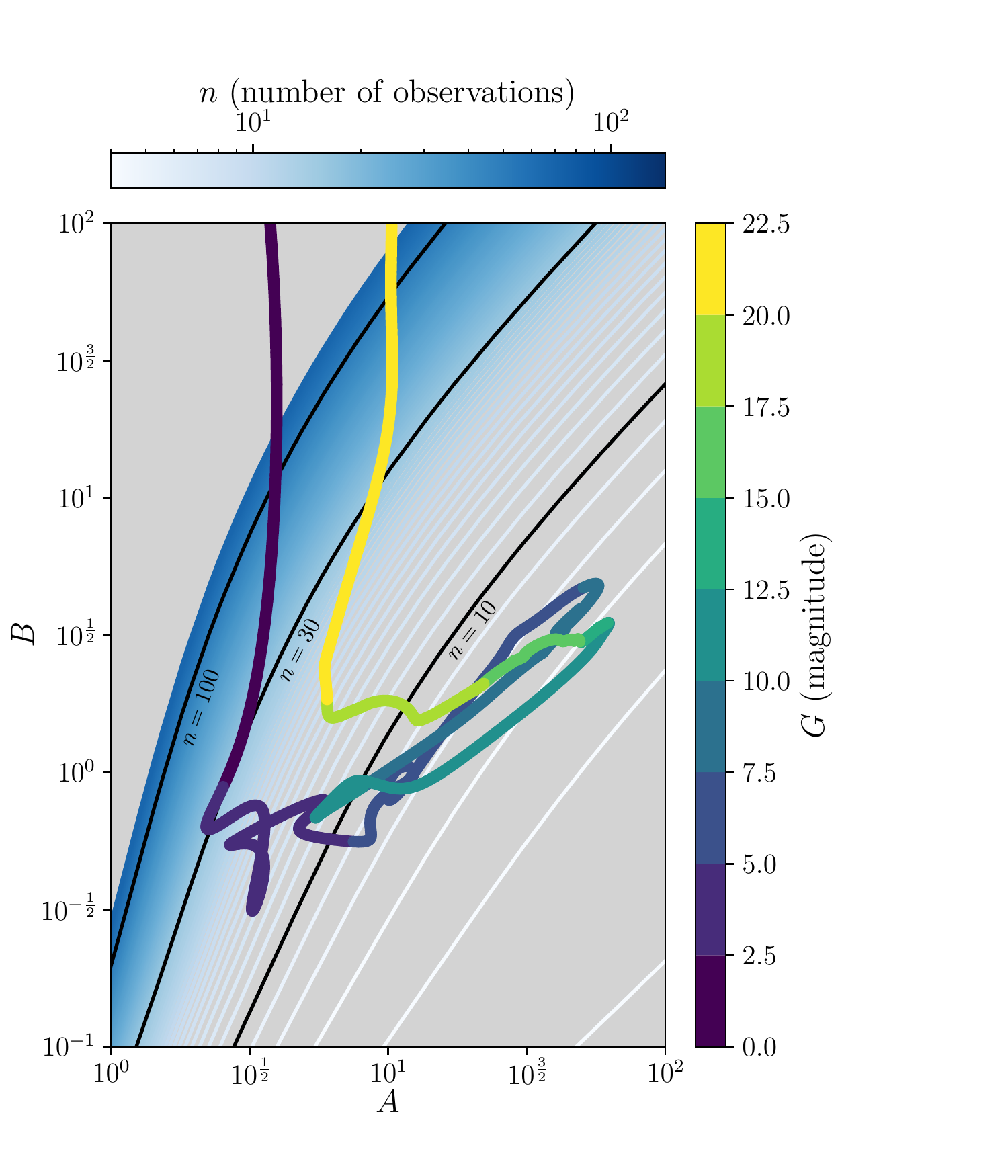}
	\caption{A schematic diagram illustrating the methodology of Appendix \ref{sec:completenessAB}. The blue lines show the contours in $(A,B)$ where our Model AB selection function would imply 99\% completeness for sources which were observed $n$ times. The multi-coloured line gives the path of our Model AB selection function (without crowding included, see Sec. \ref{sec:results}) in the space $(A,B)$ as a function of magnitude. The location of this path in $(A,B)$ relative to the contours gives the minimum number of observations that a source must receive for it to be detected with at least 99\% confidence.}
	\label{fig:schematic}
\end{figure}

Having obtained the posterior on the detection probability for Model AB, a quantity of immediate interest in the implied completeness of \gaia DR2 across the sky. The probability that a star -- with magnitude $G$ and which transited the \gaia FoV $n$ times -- will make it into \gaia DR2 is simply the expectation $\mathbb{E}_{\theta}[\operatorname{I}_{\theta}(5,n-4)|n,A,B]$, given that $\theta\sim\operatorname{Beta}(A,B)$ and where $(A,B)$ are both functions of $G$ (and perhaps also the source density $\rho$). This expectation has the following closed form solution
\begin{align}
K(n,A,B) &= \mathbb{E}_{\theta}[\operatorname{I}_{\theta}(5,n-4)|n,A,B] \nonumber \\
 &= 1-\sum_{m=0}^{4}\binom{n}{m}\frac{\operatorname{B}(A+m,B+n-m)}{\operatorname{B}(A,B)}, \label{eq:completeness}
\end{align}
which is obtained by noticing that the incomplete beta function can be expanded as a sum of products of $\theta$ and $1-\theta$. For each $n$ we identified the contour $K(n,A,B)=99\%$ in $(A,B)$-space and show these contours in blue in Fig. \ref{fig:schematic}, with the posterior median of $(A,B)$ of Model AB overlaid. Fig. \ref{fig:schematic} can be interpreted in two ways:
\begin{itemize}
	\item Suppose a source with magnitude $G$ was observed $n$ times. If $\left(A,B\right)$ lies to the right of the $n$-contour, then the source is in \gaia DR2 with greater than 99\% probability.
	\item Any source of magnitude $G$ with $n$ such that the $n$-contour is to the left of $\left(A,B\right)$ is in \gaia DR2 with greater than 99\% probability.
\end{itemize}
The dip in detection probability at $G=11$ (see Fig. \ref{fig:completeness}) appears in Fig. \ref{fig:schematic} as a blue loop, with the minimum number of observations for a location on the sky to be 99\% complete going from eight at $G=10$ to thirteen at $G=11$ and back down to eight at $G=12$. The minimum number of observations then rapidly increases, until at $G=21.5$ there are no locations on the sky that received sufficient observations to be 99\% complete.

To calculate the completeness map in the top panel of Fig. \ref{fig:completenessmaps}, we interpolated the values of $\left(A(G),B(G)\right)$ in each bin with a cubic spline and found for each $n$ the value of $G$ such that $K\left(n,A(G),B(G)\right)=99\%$. We applied the same methodology for the bottom panel of Fig. \ref{fig:completenessmaps}, except that we interpolated with a bicubic spline in both $G$ and source density $\rho$, and found for each $n$ the value of $G$ and $\rho$ such that $K\left(n,A(G,\rho),B(G,\rho)\right)=99\%$. \edits{The value of $\rho$ used in each $\textsc{nside}=4096$ HEALPix pixel was set as the mean density of \gaia sources in the parent $\textsc{nside}=1024$ HEALPix pixel.}

We note that determining the completeness of Model T corresponds to identifying the $G$ where $\operatorname{I}_{T(G)}(5,n-4)=99\%$. The inverse of the regularized incomplete beta function is a standard function in numerical libraries, making this a straightforward computation.

\section{Gating sources}
\label{sec:gating}
\edits{This appendix briefly describes the possible impact of the CCD gates on the completeness of \gaia with respect to sources with magnitudes in the range $10<G<12$, but we refer interested readers to \citet{Gaia2016} for a more in-depth discussion.}

When scanning a source, the Gaia CCDs accumulate charge under the photocells of the source image and move the charge with the source along the focal plane. By the end of the CCD panel, the charge has been integrated across the entire scan and it is this integration of charge which is used to calculate the source parameters such as the G-band magnitude and position of the photo-center.

For sources brighter than $G=12$, the CCDs saturate in charge before the end of the CCD panel is reached which is clearly problematic for determining a precise photo-center or brightness. To counteract this, Gaia performs an on-board truncation of the integral after an amount of time. If the skymapper CCD determines that the source is brighter than $G=12$, the charge will be integrated for half of the scan before truncation hence avoiding saturation. This truncation is called a 'gate configuration' \citep{Carrasco2016}. In actual fact Gaia applies multiple gate configurations at different magnitudes brighter than $G=12$ which can be seen as the discontinuous changes in the saw-tooth green line of Figure 10 of \citet{Evans2018}. The bump in photometric error at $G=11$ is caused by these gates since at each gate the flux used to infer $G$ is halved.  The effect of this on astrometric parameter estimation can also be seen in Fig. 9 of \citet{Lindegren2018} where the running median of along-scan astrometric measurement uncertainties from the raw image parameters clearly mimics the saw-tooth pattern whilst the final model uncertainty displays the same bump at $G=11$. 

Due to the uncertainty in onboard magnitude assignment, stars can be assigned to the wrong gate configurations in different scans. This incorrect assignment can lead to insufficient flux being recorded and the observation not being taken which will reduce the detection probability.

\bsp	
\label{lastpage}
\end{document}